\documentclass[10pt,aps,prd,reprint,nofootinbib,floats,floatfix,amsfonts,amssymb,amsmath,preprintnumbers,notitlepage,superscriptaddress,twocolumn]{revtex4-1}

\usepackage{url}
\usepackage{graphicx,color} % Figure support
\usepackage{hyphenat} % Allow line break on hyphenated words
\usepackage{amsmath, amssymb, amsfonts,amsbsy,amsthm} %Math and such
\usepackage{float}
\usepackage{subcaption}
\usepackage{soul,xcolor} % to bar text
\usepackage{booktabs}
\usepackage{multirow}

\setstcolor{red}

\newcommand{\orcid}[1]{\href{https://orcid.org/#1}{\includegraphics[scale=0.15]{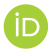}}}
\newcommand{\Kappa}[0]{\scalebox{1.5}{$\kappa$}}

\usepackage[normalem]{ulem}

\begin{document}

\title{Coherent injection of magnetic noise and its impact on gravitational-wave searches}

\author{Kamiel Janssens\orcid{0000-0001-8760-4429}}
\affiliation{Universiteit Antwerpen, Prinsstraat 13, 2000 Antwerpen, Belgium}
\affiliation{Universit\'e C$\hat{o}$te d’Azur, Observatoire de la C$\hat{o}$te d’Azur, CNRS, Artemis, 06304 Nice, France}
\affiliation{Department of Physics, The University of Adelaide, Adelaide, SA 5005, Australia}
\affiliation{ARC Centre of Excellence for Dark Matter Particle Physics, Melbourne, Australia}

\author{Jessica Lawrence\orcid{0000-0003-1222-0433}}
\affiliation{Department of Physics and Astronomy, Texas Tech University, Lubbock, TX 79409, USA}

\author{Anamaria Effler\orcid{https://orcid.org/0000-0001-8242-3944}}
\affiliation{LIGO Livingston Observatory, Livingston, LA 70754, USA}

 \author{Robert M. S. Schofield}
 \affiliation{University of Oregon, Eugene, OR 97403, USA}

\author{Max Lalleman\orcid{0000-0002-2254-010X}}
\affiliation{Universiteit Antwerpen, Prinsstraat 13, 2000 Antwerpen, Belgium}

\author{Joseph Betzwieser\orcid{0000-0003-1533-9229}}
\affiliation{LIGO Livingston Observatory, Livingston, LA 70754, USA}

\author{Nelson Christensen \orcid{0000-0002-6870-4202}}
\affiliation{Universit\'e C$\hat{o}$te d’Azur, Observatoire de la C$\hat{o}$te d’Azur, CNRS, Artemis, 06304 Nice, France}
\affiliation{Physics and Astronomy, Carleton College, Northfield, MN 55057, USA}

\author{Michael W. Coughlin\orcid{0000-0002-8262-2924}}
\affiliation{School of Physics and Astronomy, University of Minnesota, Minneapolis, MN 55455, USA}

\author{Jennifer C. Driggers\orcid{0000-0002-6134-7628}}
\affiliation{LIGO Hanford Observatory, Richland, WA 99354, USA}

\author{Adrian F. Helmling-Cornell\orcid{https://orcid.org/0000-0002-7709-8638}}
\affiliation{University of Oregon, Eugene, OR 97403, USA}

\author{Timothy J. O'Hanlon\orcid{0009-0007-6209-4654}}
\affiliation{LIGO Livingston Observatory, Livingston, LA 70754, USA}

\author{Eric A. Quintero\orcid{0000-0002-4269-3445}}
\affiliation{Massachusetts Institute of Technology, Cambridge, MA 02139, USA}

\author{Juliedson A. M. Reis\orcid{0000-0001-7372-1827}}
\affiliation{National Institute for Space Research, São José dos Campos, SP 12227-010, Brazil}

\author{Nick van Remortel\orcid{https://orcid.org/0000-0003-4180-8199}}
\affiliation{Universiteit Antwerpen, Prinsstraat 13, 2000 Antwerpen, Belgium}

\date{\today}

\begin{abstract}
Correlated noise sources, particularly magnetic noise, form a risk to future gravitational-wave searches aimed at detecting the gravitational-wave background. Potential noise contamination is investigated by making noise projections which typically rely on an accurate measurement of the coupling strength of the noise to the detector. To make these projections, we inject, for the first time, broadband, coherent magnetic noise between two gravitational-wave detectors, LIGO Hanford and LIGO Livingston, separated by several thousands of kilometers.
We describe the noise injection as well as its impact on the analysis pipelines and investigate the accuracy of noise projection techniques used in the past decade. Finally, we present a proof-of-concept demonstration of noise subtraction using Wiener filtering, while also highlighting potential risks associated with this method. This unique data set with correlated noise caused by magnetic field fluctuations in two gravitational-wave detectors, as well as in an array of witness sensors, provides an excellent testing ground for additional future studies. Ultimately, this study demonstrates that Wiener filtering is effective and can be applied in the eventual detection of the gravitational-wave background by the LIGO-Virgo-KAGRA Collaboration.
\end{abstract}

\maketitle

%%%%%%%%%%%
\section{Introduction}
\label{sec:Introduction}
%%%%%%%%%%%

The observation of a gravitational-wave (GW) background (GWB) with Earth-based interferometric detectors LIGO \cite{ADV_LIGO_2015}, Virgo \cite{VIRGO:2014yos} and KAGRA \cite{PhysRevD.88.043007} should provide new insights into the population of stellar-mass black holes or, even more interestingly, cosmological processes in the early universe~\cite{Christensen_2018}. However, previous research has shown that such a search could be affected by correlated noise \cite{Thrane:2013npa,Thrane:2014yza,Coughlin:2016vor,PhysRevD.96.022004,Coughlin:2018tjc,PhysRevD.100.082001,Meyers:2020qrb,PhysRevD.104.122006,PhysRevD.107.022004,PhysRevD.106.042008}. More specifically, Schumann resonances  \cite{Schumann1,Schumann2} can impact searches for an isotropic GWB using a network of detectors separated by Earth-scale distances. More recently, the effect on directional searches for a GWB was also investigated \cite{PhysRevD.111.082005}. Schumann resonances are electromagnetic resonances of the spherical cavity bounded by the ionosphere and the Earth's surface, excited by lightning strikes. These magnetic field fluctuations can couple to GW detectors in multiple ways, e.g. by direct coupling to permanent magnets in actuation systems or by acting upon electronics and cabling \cite{Thrane:2013npa,Thrane:2014yza,galaxies8040082,Nguyen_2021,soni2024ligodetectorcharacterizationhalf}. The fundamental mode of Schumann resonances occurs around 7.8 Hz.

A recent study notes that Schumann resonances could potentially limit the search for an isotropic GWB up to several tens of Hz when LIGO and Virgo reach their A+ and AdV+ design sensitivities, respectively \cite{PhysRevD.107.022004}. The effect of lightning-induced magnetic fields for future detectors such as the Einstein Telescope \cite{Punturo_2010,Hild:2010id,Maggiore:2019uih,Amann:2020jgo,ETdesignRep} and Cosmic Explorer \cite{Reitze2019Cosmic} is expected to be even more severe if one does not try to reduce the magnetic coupling for these future projects \cite{PhysRevD.104.122006,PhysRevD.107.022004}.  
Lightning and Schumann resonances could also affect the search for short duration GW transients~\cite{Kowalska_Leszczynska_2017,PhysRevD.107.022004}.
Over the last decade a large amount of effort has focused on characterizing the effects of these noise sources as well as on possible mitigation techniques. However, all of these efforts rely either on producing projections or analysing simulated data. 
We performed a coherent injection of magnetic noise between the LIGO observatories and various analyses of the corresponding data to gain a more holistic understanding of how correlated magnetic noise affects GW detector sensitivity to the GWB.

In Sec. \ref{sec:Injection} we describe the infrastructure used to create coherent magnetic fields in the central stations of the LIGO Hanford observatory (LHO) and LIGO Livingston observatory  (LLO). Furthermore, we discuss the creation of the signal which was injected as well as its recovery. 
Afterwards we construct a noise projection on the detectors' amplitude spectral density (ASD) in Sec. \ref{sec:ASDBudget}. In Sec. \ref{sec:GWBBudget} we project the effect of the coherent magnetic noise on the search for an isotropic GWB, which is in line with earlier work \cite{PhysRevD.104.022004,PhysRevD.107.022004}. In Sec. \ref{sec:Isotropic} we look into how this correlated noise would show up in a real GWB analysis. For this purpose we use the {\tt pygwb}-pipeline \cite{Renzini:2023qtj} used by the LIGO, Virgo and KAGRA (LVK) Collaboration to search for an isotropic GWB. 
In Sec.~\ref{sec:NoiseSubtraction} we use the data from the correlated noise injection as a case study for applying Wiener filtering noise subtraction to similar datasets. Whereas future, realistic scenarios might be different since the correlated magnetic signals will be significantly lower in amplitude, our injections and analyses represent a good proof-of-concept on which future work can build.
Finally, in Sec.~\ref{sec:Conclusion} we conclude by listing the key take-away messages from this first coherent magnetic noise injection between two GW detectors separated by thousands of kilometers. Furthermore, we provide suggestions for future work to address some of the shortcomings of our study.

%%%%%%%%%%%%%
\section{Correlated injection}
\label{sec:Injection}
%%%%%%%%%%%%%

The LIGO detectors have several large wire coils in their experimental buildings. These are used to measure the coupling strength of magnetic fields to the GW strain measurement by generating magnetic fields several orders of magnitude larger than the ambient magnetic field.
In this work we will use these injection coils to generate a correlated injection between LHO  and LLO.

\subsection{Injected magnetic signal}

\begin{figure}
\centering
\includegraphics[width=0.98\linewidth]{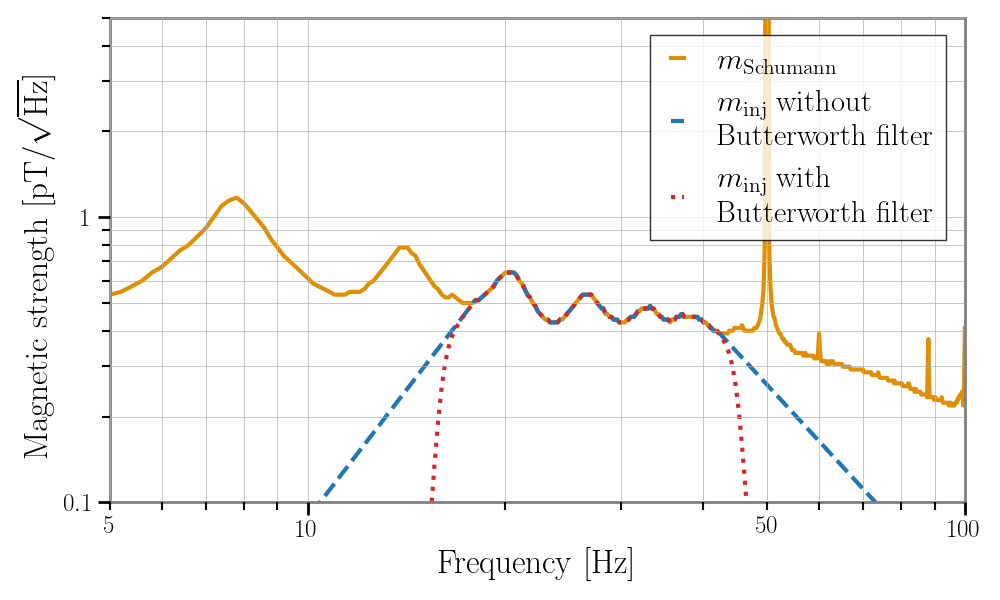}
\caption[Proposed Schumann resonance spectrum for correlated magnetic injection.]{The magnetic spectrum, $m_{\mathrm{inj}}$, that will be used for coherent injections between multiple GW detectors to test how correlated noise affects GWB pipelines.
The spectrum is compared to the Schumann resonance spectrum observed at Sos Enattos (10\% scaled to 95\% amplitude, see Fig. 1 of \cite{PhysRevD.104.122006}). In addition, $m_{\mathrm{inj}}$ is shown both before and after applying the 32nd order low- and high-pass Butterworth filters.}
\label{fig:Minjection}
\end{figure}

To illustrate the ability to create a synchronized magnetic injection between LHO and LLO, a first proof-of-concept injection was performed on June 28 2023. A white noise, broadband spectrum was injected for approximately 5 minutes between 10Hz and 40Hz \cite{alog:O4CorrTestInjLHO,alog:O4CorrTestInjLLO}. 
Its successful completion enabled us to plan the injection of a $\sim$ 45min long, physically-motivated magnetic noise spectrum to test our pipelines for the effect of correlated magnetic noise.
For this injection we will focus on the frequency band between $\sim$20Hz and $\sim$40Hz, i.e. spanning the third to sixth order Schumann resonances. 
The injected spectrum will be based on the one observed at the Sos Enattos mine, presented in Fig. 1 of \cite{PhysRevD.104.122006}. We will use the 10\% percentile spectrum scaled to the 95\% percentile amplitude as presented in dark orange in \cite{PhysRevD.104.122006}. To not overburden the magnetic injection system, the injected spectrum will be tapered off outside the target frequency band, i.e. below 20Hz and above 40Hz. The search for an isotropic GWB typically starts analyzing data above 20Hz, thus there is no interest from the analysis point of view to also inject magnetic fields at lower frequencies\footnote{Note that in a realistic physical scenario these lower frequency components will be present and (can) leave a visible imprint on the observed strain cross spectral density.}.
Above 40Hz, the magnetic coupling will have decreased significantly such that little to no   effect from the injection is expected, given a realistic injection amplitude~\cite{PhysRevD.104.022004}. This was also the case for the test injection demonstrated earlier \cite{alog:O4CorrTestInjLHO,alog:O4CorrTestInjLLO}. Therefore we want to focus the available power of the amplifier in the most relevant frequency region, without diluting the injection by spreading over too large of a frequency band. More concretely, the injected magnetic $m_{\mathrm{inj}}$ spectrum will have the following frequency content:
\begin{equation}
    \label{eq:corrmaginjspectrum}
    \begin{aligned}
        &  f  < \mathrm{18Hz}  : 
           m_{\mathrm{inj}} = m_{\mathrm{Schumann}}(\mathrm{18Hz}) \times \left(\frac{f}{\mathrm{18Hz}}\right)^{3} \\
         \mathrm{18Hz} \leq\ & f \leq \mathrm{42Hz}  : 
          m_{\mathrm{inj}} = m_{\mathrm{Schumann}}(f) \\
         \mathrm{42Hz} <\ & f    : 
          m_{\mathrm{inj}} = m_{\mathrm{Schumann}}(\mathrm{42Hz}) \times \left(\frac{f}{\mathrm{42Hz}}\right)^{-2.5},        
    \end{aligned}
\end{equation}
where $ m_{\mathrm{Schumann}}$ is the magnetic spectrum observed at Sos Enattos. The slopes of the power-law tapering are arbitrarily chosen to create a somewhat smooth transition. Afterwards this spectrum is multiplied by a 32nd order high-pass Butterworth filter with a cutoff frequency of 16Hz and a 32nd order low-pass Butterworth filter with a cutoff frequency of 42Hz.
These different spectra are shown in Fig. \ref{fig:Minjection}.

Based on the earlier test injection, the amplitude of the final injection was chosen to be $\mathcal{O}(10^4)$ times the amplitude of the environmental Schumann spectrum shown in Fig. \ref{fig:Minjection} near the `Vertex' magnetometers.

\begin{figure*}
\centering
\includegraphics[width=0.98\linewidth]{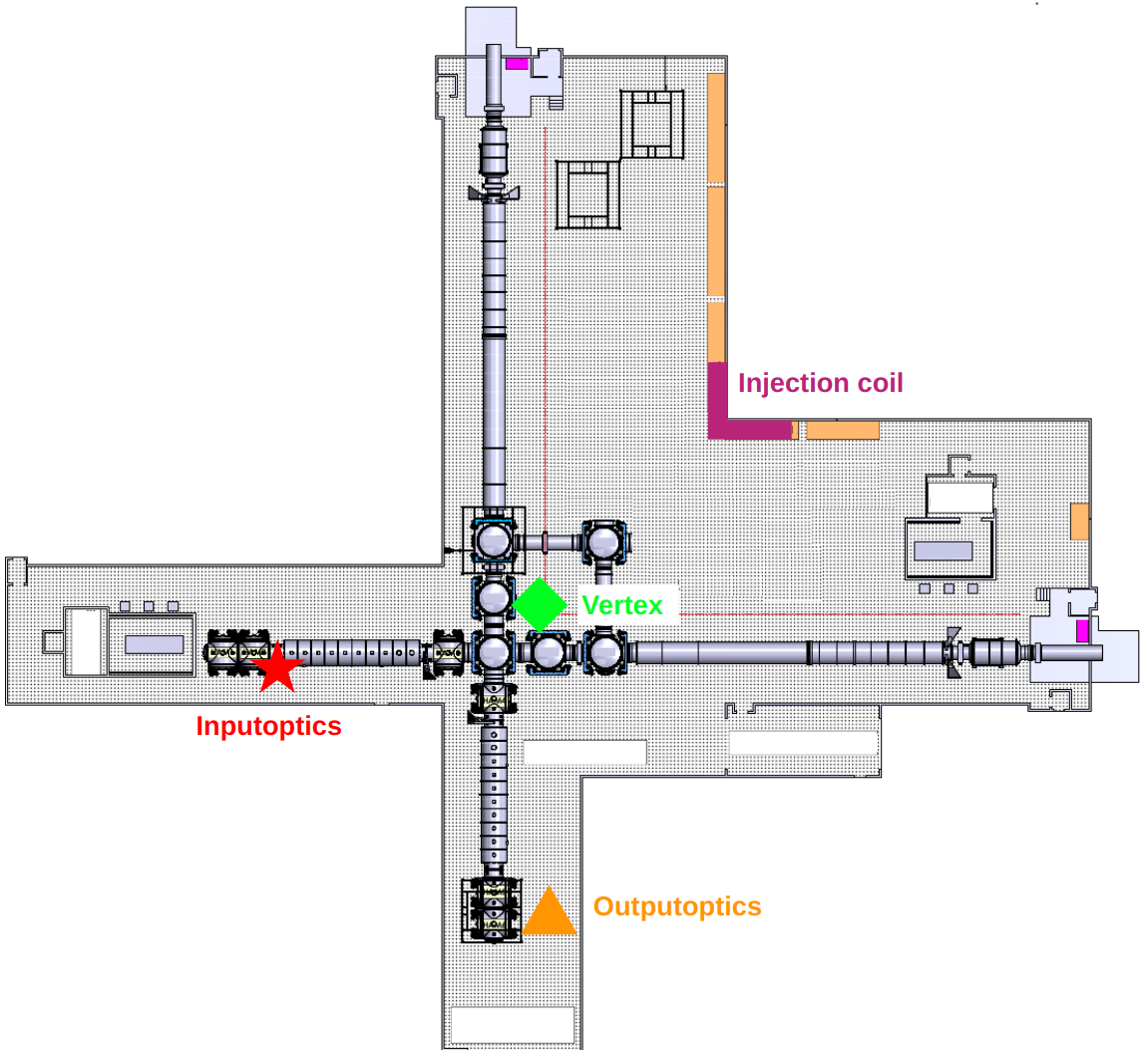}
\caption{Map of the LHO central station (CS) showing the positions of the magnetometers, marked as follows: green diamond for `Vertex', red star for `Inputoptics', and orange triangle for the `Outputoptics'. The L-shaped purple box represents the location of the large injection coil on the wall used in this work.}
\label{fig:LHO_map}
\end{figure*}

Figure \ref{fig:LHO_map} shows a map of LHO's central station (CS) with the location of the different magnetometers indicated with a green diamond (`Vertex'), red star (`Inputoptics') and orange triangle (`Outputoptics'). The L-shaped purple box indicates the location of the injection coil used in this study. LLO has a very similar layout considering the locations of our sensors and the coil, hence we do not show a separate map. Here we would like to highlight that the coil has a rectangular shape and is attached to this corner of the wall. This implies the coil has an angle of 90$^{\circ}$ in its middle.

\subsection{Recovered signal}

On Dec 20 2023, a $\sim$ 45min long synchronous, magnetic injection was performed at LHO and LLO \cite{alog:O4CorrInjLHO,alog:O4CorrInjLLO}.
Whereas the natural Schumann resonances are expected to be highly isotropic and homogeneous across the entire detector site and buildings, our amplified injected Schumann signal does not exhibit these characteristics. Due to the nature of the placement of the injection coils there will be both a directionality to the signal as well as large fluctuations in signal strength across the building. For this reason we will look at the observed magnetic field in several locations throughout the central station and project their contribution to the strain ASD.
More specifically we will be analyzing data from magnetometers near the `Vertex', `Inputoptics' and `Outputoptics' for both LHO and LLO. In the main text we will discuss the `Vertex' location as there are minimal differences between the different locations. For completeness, we present additional results from the other witness sensors in Appendix \ref{appendix:inputOutput}. To ensure no features are introduced due to the sudden switch on and off of the injection, we analyze 43 min of data starting at Dec 20 - 22:27:18  UTC for the injection. As background estimates we use 43 min of data starting at Dec 20 - 17:01:00 UTC and Dec 21 - 02:01:00 UTC as reference for before and after the injection, respectively. These times are not directly adjacent nor part of the same detector lock period. The injection was started shortly after acquiring lock and one of the detectors lost lock shortly after finalizing the injection.

\begin{figure*}
    \centering
    \includegraphics[width=0.49\textwidth]{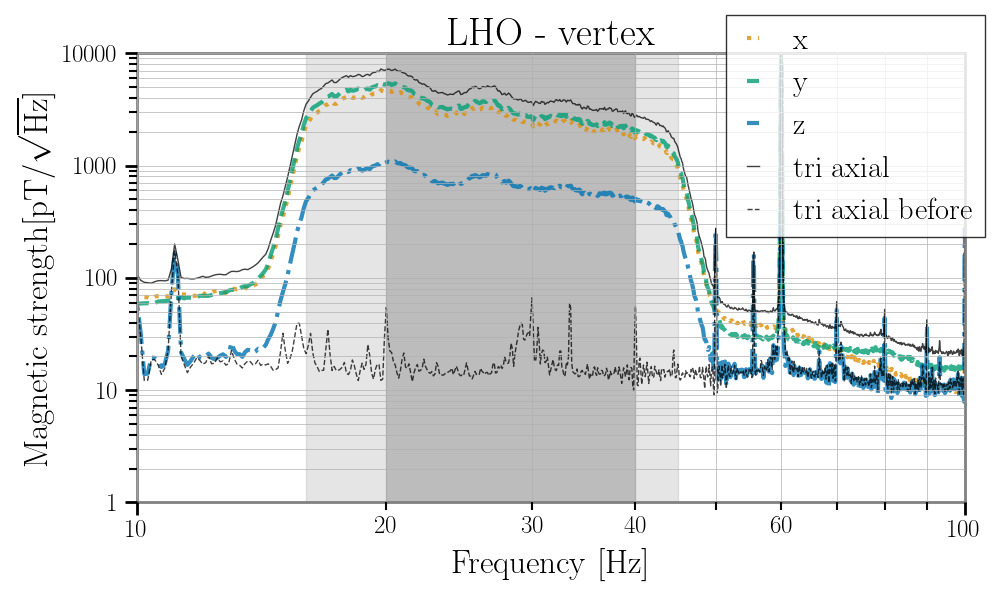}
    \includegraphics[width=0.49\textwidth]{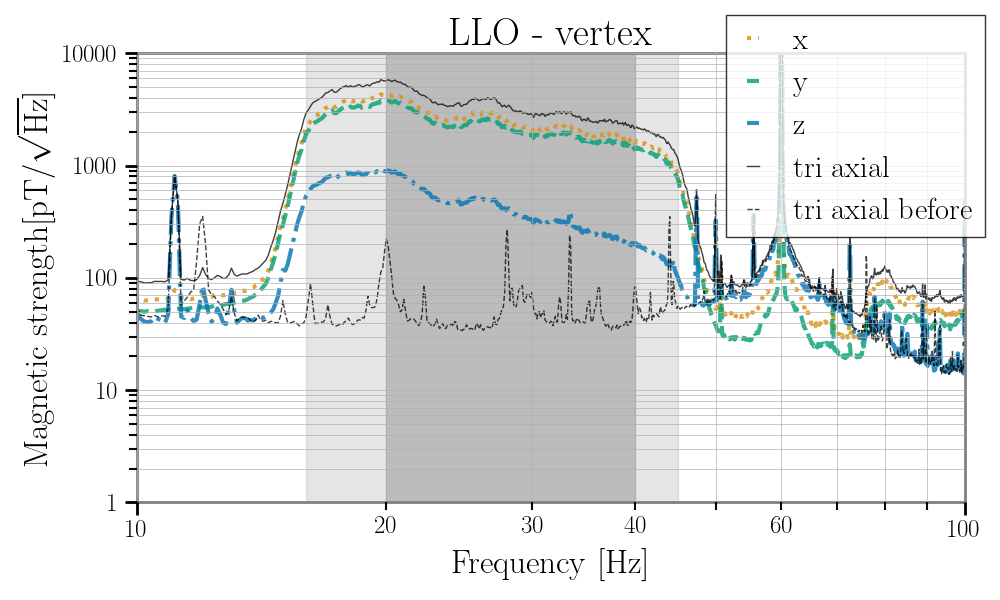}
    \caption{The magnetic amplitude spectral density for LHO (left) and LLO (right) observed during the injection at the central station (CS) `Vertex' magnetometer. For the time of the injection we show the three independent components, x (dotted, yellow), y (dashed, green), z (dot-dashed, blue) as well as tri-axial (solid, black). For comparison we also show the magnetic noise observed during the reference time before the injection (dashed, black). The dark and light gray vertical bands show the frequency range in which the noise was injected and tapered off, respectively.}
    \label{fig:MagInj_ASD}
\end{figure*}

Figure \ref{fig:MagInj_ASD} shows the observed ASD of the $ x$, $y$, and $z$ magnetic fields for LHO (left panel) and LLO (right panel), along with the quadrature sum of the components, labeled `tri-axial,' both before and after the injection. The $x$ and $y$ directions are defined along the interferometer arms, while the $z$ direction is normal to the Earth's surface. We also refer to the magnetic measurements along the $i= x,y,z$ direction of the Hanford (Livingston) detector by the notation $H_i$ ($L_i$) or e.g. Hx (Lx).

\begin{figure*}
    \centering
    \includegraphics[width=0.49\textwidth]{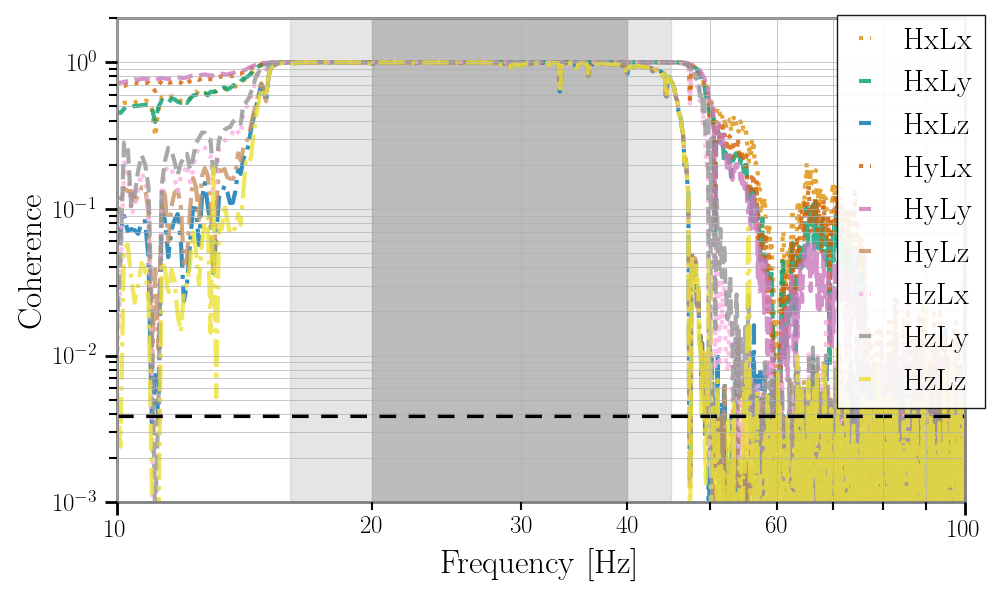}
    \includegraphics[width=0.49\textwidth]{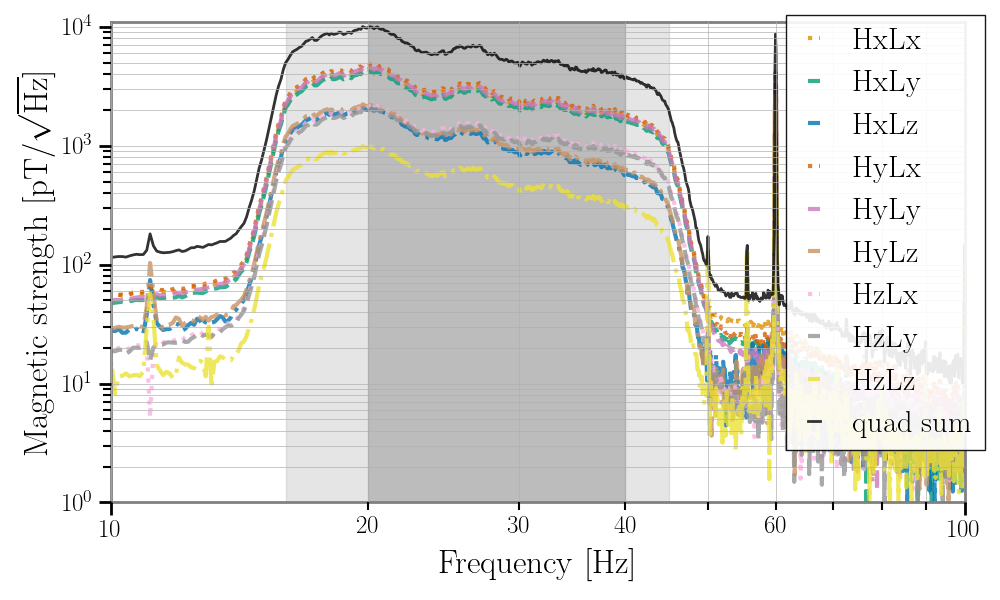}
    \caption{ The magnetic coherence (left) and cross spectral density (right) between the `Vertex' magnetometers at LHO and LLO during the injection. We present all nine combinations for two tri-axial measurements of the magnetic field. The dark and light gray vertical bands show the frequency range in which the noise was injected  and tapered off,  respectively.}
    \label{fig:MagInj_CohCSD}
\end{figure*}

Figure \ref{fig:MagInj_CohCSD} presents the observed coherence and cross spectral density (CSD) between the different orientations of the `Vertex' magnetometers. Within the region of the injection (grey bands) we have very high coherence, close to unity across the entire frequency band for all nine direction pairs.
Based on the right panel of Fig. \ref{fig:MagInj_CohCSD} we can conclude that the amplitude of a CSD including vertical magnetic fields ($z$-direction) is smaller than their horizontal counterparts. This is in line with Fig. \ref{fig:MagInj_ASD}, as well as  expectations since the injection coils predominantly create horizontal magnetic field fluctuations.

In the last part of this section we will estimate what the proportional strength of the magnetic field is at a given distence $d$ from a circular coil to compare this to the difference in strength we observe at our different magnetometers. We would like to stress that this is a highly idealized calculation as in reality our coils or not circular but square and on top of that or not contained to a single plain but have a 90$^{\circ}$ degree in its middle. Neverthless we believe this calculation can provide an order of magnitude indication of the expect ratio in strengths.
The expected magnetic field at a distance $d$ from the coil is given by\footnote{Note, here we have neglected any width of the coil and the connecting line between the observer and the center of the coil is perpendicular to the plane of the coil, i.e. the observer has a `central' placement with respect to the coil.} \cite{GriffithsEDM},
\begin{equation}
    \label{eq:MagFieldCoil}
    B(I,d) = \frac{\mu_0 N I}{2} \frac{r^2}{(r^2+d^2)^{3/2}},
\end{equation}
where $\mu_0$ is the magnetic permeability of the vacuum, $N$ the number of turns of the coil, $r$ the radius of the coil and $I$ the current flowing through the coil.
If we neglect the radius of the coil (several meters) with respect to the distances ($\sim$ 20m for `Vertex' and $\sim$ 40m for `Inputoptics' and `Outputoptics') we estimate the magnetic field to be about 8 times larger at the `Vertex' magnetometer. This roughly matches what we observe, indicating the observed fields match our expectations.

We conclude this section by confirming that this injection successfully created coherent magnetic fields between magnetometers separated by $\sim$3000~km. In the next section we will investigate whether the injection was strong enough to be observed in the GW-sensitive strain channel.

%%%%%%%%%%%%%
\section{ASD Magnetic budget}
\label{sec:ASDBudget}
%%%%%%%%%%%%%

We can estimate the effect caused by magnetic fields on the strain measurement of interferometer $I$ ($S_{mag,I}$) as the product of the magnetic ASD ($\Tilde{m}_I(f)$) with the magnetic coupling function ($\Kappa_I(f)$), i.e.: $S_{mag,I} = \Kappa_I(f) \times  \Tilde{m}_I(f)$. The magnetic ASD was described in the previous section and shown in Fig. \ref{fig:MagInj_ASD}.
For the values of the magnetic coupling we use the magnetic coupling specific for each of the three locations under investigation (`Vertex', `Inputoptics' and `Outputoptics') as measured during the observing run O4a when this injection was performed \cite{2024CQGra..41n5003H,soni2024ligodetectorcharacterizationhalf}. Please note: this is an independent injection from the one performed and under investigation for this paper.
For the error on the magnetic budget we use the intrinsic error which is estimated to be a factor 2, due to the innate difficulties in measuring the accurate coupling as described in \cite{Nguyen_2021}.

\begin{figure*}
    \centering
    \includegraphics[width=0.49\textwidth]{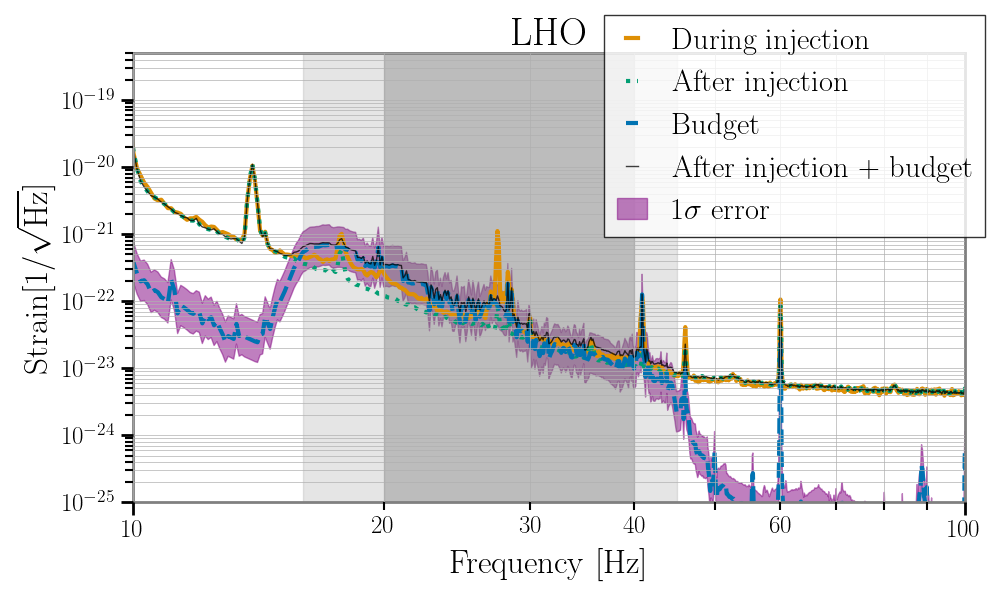}
    \includegraphics[width=0.49\textwidth]{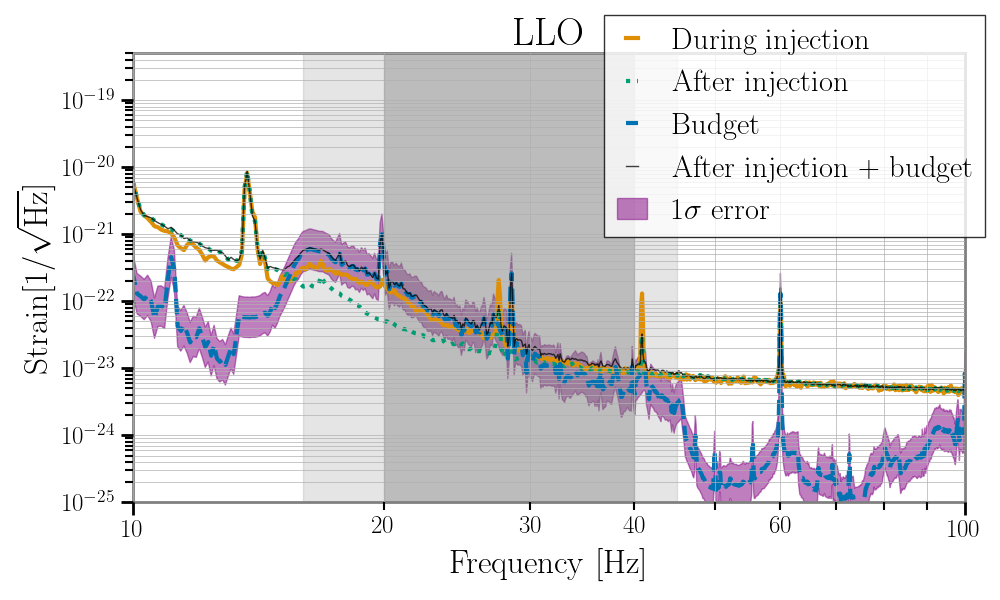}
    \caption{Projection of the magnetic noise on the GW strain ASD of LHO (left) and LLO (right) during the injection. The dark and light gray vertical bands show the frequency range in which the noise was inject and tapered off, respectively.}
    \label{fig:MagInj_ASDBudgets}
\end{figure*}

In Fig.~\ref{fig:MagInj_ASDBudgets} we present the magnetic noise projection for LHO (left panel) and LLO (right panel), using the observed magnetic noise at the `Vertex' location. Note, here we have used the quadrature sum of the x-, y- and z-directions to use the tri-axial magnetic noise estimate in this projection. In Appendix \ref{appendix:inputOutput} the equivalent noise projections are shown for the `Inputoptics' and `Outputoptics' locations. Apart from the `Outputoptics' at LHO, we do observe significant and similarly strong coupling projections based on the different locations at both LHO and LLO.
However, as can be seen in Fig. \ref{fig:MagInj_ASDBudgets} the projected effect is marginally larger than the observed effect. Nevertheless they agree with each other within the 1$\sigma$ uncertainty on the magnetic coupling function. 
However, if the three locations `Vertex' and `Inputoptics' and `Outputoptics' (presented in Appendix \ref{appendix:inputOutput}) would be considered as separate couplings the overestimation would be larger.

\begin{figure}
    \centering
    \includegraphics[width=0.49\textwidth]{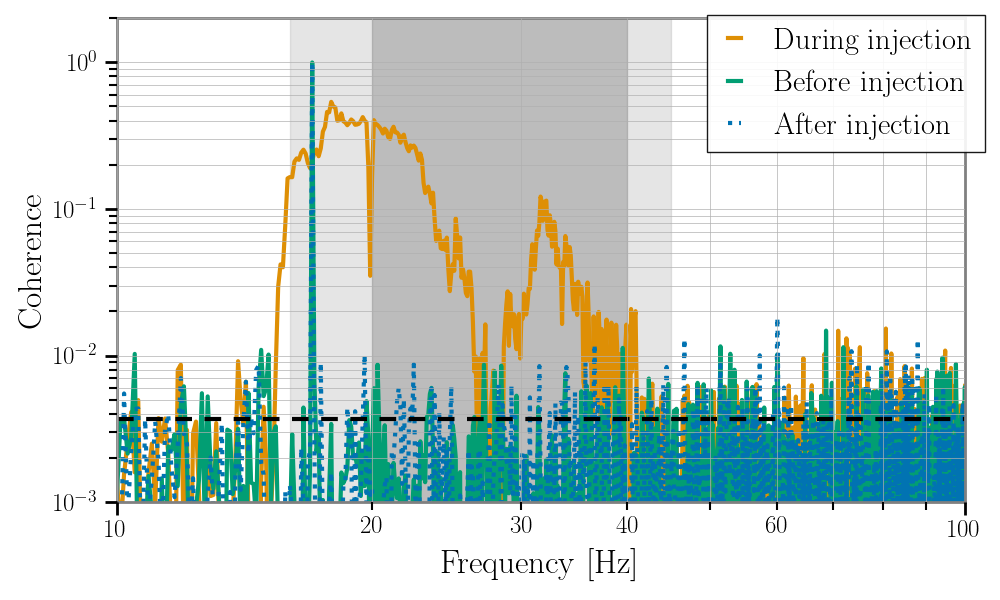}
    \caption{Magnetic coherence between the GW strain of LHO and LLO during the injection. The dark and light gray vertical bands show the frequency range in which the noise was injected and tapered off, respectively.}
    \label{fig:MagInj_StrainCoherence}
\end{figure}

Finally, we want to understand to what  extent the correlated magnetic noise, which couples significantly to the strain channel,  also causes coherent effects between the observed strain at LHO and LLO. The magnetic coupling functions which are measured at the sites focus only on the amplitude and not on the phase. Therefore it is possible that  a site-specific phase shift is introduced when magnetic fields couple to the detectors strain sensitivity. As shown in Fig. \ref{fig:MagInj_StrainCoherence} we  observe significant strain-strain coherence between $\sim$16Hz and $\sim$40Hz. 
Features around 28Hz and 41Hz are indicative of suspensions resonances, most likely from the signal recycling optics. These are excited by the magnetic injection pushing on the suspension magnets, but are not fully captured in the magnetic coupling projection.

We conclude this section by confirming the amplitude of the injected magnetic field fluctuations was such that not only was coherence observed in the magnetometers, but it was also observed  in the GW-sensitive strain channel for two widely separated ($\sim$3000~km) interferometeric GW detectors. This provides the ideal testing ground to further investigate the effect of correlated, cross-site, broadband noise effects. In the following sections we will look into these effects for the search for an isotropic GWB.

%%%%%%%%%%%%%
\section{GWB Magnetic budget}
\label{sec:GWBBudget}
%%%%%%%%%%%%%

The isotropic GWB analysis tries to measure $\Omega_{\mathrm{GW}}$\cite{PhysRevD.46.5250,PhysRevD.59.102001,Romano:2016dpx}, the energy density $\text{d}\rho_{\mathrm{GW}}$ contained in a logarithmic frequency interval $\text{d}lnf$, divided by the critical energy density $\rho_c$\footnote{Where $\rho_{\mathrm{c}} = 3H_0^2c^2/(8\pi G)$.} needed for a flat universe:
\begin{equation}
    \label{eq:omegaGW}
    \Omega_{\mathrm{GW}} = \frac{1}{\rho_c} \frac{\text{d}\rho_{\mathrm{GW}}}{\text{d}lnf}.
\end{equation}
In the absence of correlated noise, we can construct the following unbiased estimator of $\Omega_{\mathrm{GW}}$ \cite{PhysRevD.59.102001,Romano:2016dpx}
\begin{equation}
    \label{eq:Cij}
    \hat{C}_{IJ}(f) = \frac{2}{T} \frac{Re[\Tilde{s}^*_I(f)\Tilde{s}_J(f)]}{\gamma_{IJ}(f)S_0(f)},
\end{equation}
for detectors $I$ and $J$.
An explicit expression for the contribution of the correlated noise that biases this estimator can be found in Eq. 12 of \cite{Meyers:2020qrb}. 
$\Tilde{s}_I(f)$ is the Fourier transform of the time domain data $s_I(t)$ measured by interferometer $I$, $\gamma_{IJ}$ the normalized overlap reduction function~\cite{PhysRevD.46.5250} which encodes the baseline's geometry, and $S_0(f)$ is given by $S_0(f)=(3H_0^2)/(10\pi^2f^3)$, where $H_0$ is the Hubble-Lema\^\i tre constant. $T$ is the total observation time of the data taking period. However, if one analyses the data in separate segments, $T$ becomes the duration of the time segments used.

In the weak-signal limit for the GWB\footnote{Please note that due to the loudness of our injected signal the weak-signal limit will not hold. For this reason we will among others use a different method to estimate to estimate the unbiased uncertainty which will be explained later in this paper.}, the uncertainty on $\hat{C}_{IJ}(f)$ is given by \cite{PhysRevD.59.102001,Romano:2016dpx}
\begin{equation}
    \label{eq:sigmaGWB}
    \sigma_{IJ}(f) \approx \sqrt{\frac{1}{2T\Delta f}\frac{P_I(f)P_J(f)}{\gamma_{IJ}^2(f)S_0^2(f)}},
\end{equation}
where $\Delta f$ is the frequency resolution and $P_I(f)$ the strain PSD of detector $I$. Similar to the cross-correlation statistic (Eq. \ref{eq:Cij}) one can construct a magnetic cross-correlation statistic given by  \cite{Thrane:2013npa,Thrane:2014yza}
\begin{equation}
    \label{eq:C_Mag}
        \hat{C}_{mag,IJ}(f) = |\Kappa_I(f)||\Kappa_J(f)|  \frac{2}{T} \frac{Re[\Tilde{m}^*_I(f)\Tilde{m}_J(f)]}{\gamma_{IJ}(f)S_0(f)},
\end{equation}
where $\Kappa_I(f)$ describes the coupling of the magnetic fields to interferometer $I$ and $\Tilde{m}_I(f)$ is the Fourier transform of the time domain data $m_I(t)$ measured by a magnetometer at site $I$. $T$ is the duration of the segments used when Fourier transforming the magnetic data. $\frac{2}{T} Re[\Tilde{m}^*_I(f)\Tilde{m}_J(f)]$ is the real part of the magnetic CSD observed between detectors $I$ and $J$. In past work \cite{PhysRevD.107.022004}, the absolute value as well as the quadrature sum of the different orientations of the magnetic field fluctuations were used,
\begin{equation}
\label{eq:OmniDirecCSD}
\begin{aligned}
                CSD_{IJ} =& \frac{2}{T} \left[ \right.  |\Tilde{m}^*_{I_1}(f)\Tilde{m}_{J_1}(f)| ^2 + |\Tilde{m}^*_{I_1}(f)\Tilde{m}_{J_2}(f)|^2  \\
             &+ |\Tilde{m}^*_{I_2}(f)\Tilde{m}_{J_1}(f)| ^2 +  |\Tilde{m}^*_{I_2}(f)\Tilde{m}_{J_2}(f)|^2 \left.\right]^{1/2},        
\end{aligned}
\end{equation}
where the indices $I_1$ and $I_2$ refer, respectively, to the first and second sensors located at detectror $I$ (and similarly for detector $J$). Eq. \ref{eq:OmniDirecCSD} gives the quadrature sum for a horizontal magnetic field, since the vertical components are negligible for Schumann resonances. However, this is not entirely the case for the injections made, hence we choose to extend this equation to also include the vertical component.

\begin{figure}
    \centering
    \includegraphics[width=0.49\textwidth]{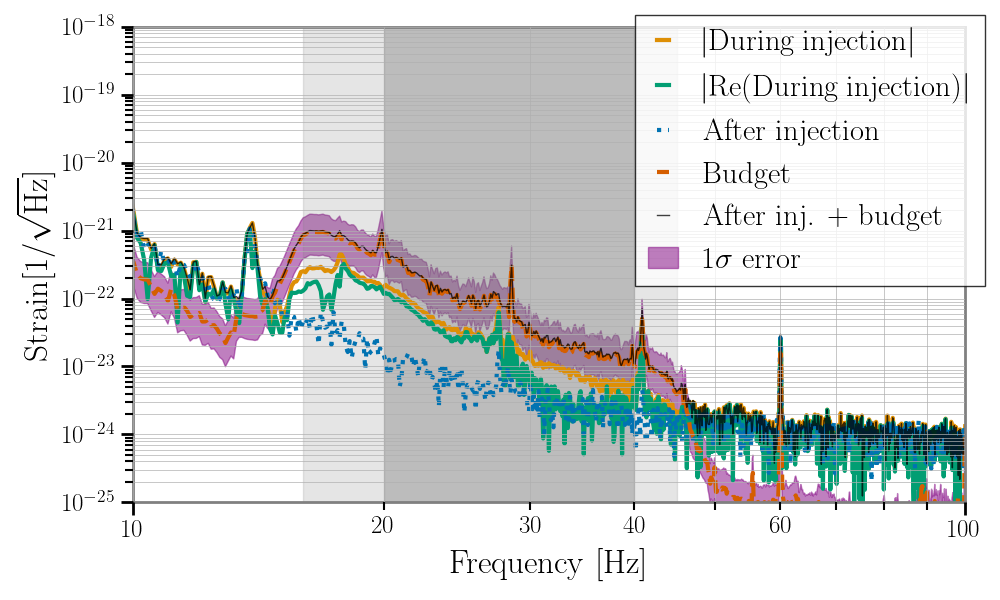}
    \caption{Correlated magnetic noise projection (orange, 1$\sigma$ error in purple) using the quadrature sum of the nine different magnetometer combinations between LHO and LLO is shown with respect to the absolute value of observed strain during the injection (yellow) and the absolute value of the real part of observed strain during the injection (green). Also shown is the the observed strain after the injection (dashed, blue) and the sum of the budget and the observed strain after the injection (black).} 
    \label{fig:MagInj_CSDBudget}
\end{figure}

In Fig. \ref{fig:MagInj_CSDBudget} we compare the magnetic noise projection with the quadrature sum of the strain CSD. The latter equals the numerator of Eq. \ref{eq:C_Mag} with the additional change that we do not use just the real part but the absolute value of the magnetic noise correlations, in order to be conservative in line with earlier work \cite{PhysRevD.107.022004}. 
Over the entire region of interest there is a discrepancy $>1\sigma$ between the observed and projected effect. If one uses only one of the main coherent components (i.e. $H_x$-$L_x$, $H_x$-$L_y$, $H_y$-$L_x$ or $H_y$-$L_y$) the budget would go down by about a factor two, in line with the ratios of the magnetic noise as shown in Fig. \ref{fig:MagInj_CohCSD}. 
This leads us to conclude that whereas the quadrature sum does indeed provide a conservative estimate, which was one of the reasons why it was introduced \cite{PhysRevD.107.022004}, it might overestimate the true noise contributions.
Secondly, we also show the absolute value of the real part of the strain-strain CSD in green in Fig. \ref{fig:MagInj_CSDBudget}. Here we notice that the correlated power starts to decrease just before 30Hz and predicts lower contributions between 30Hz and 40Hz compared to the absolute value of the strain-strain CSD. However, a similar behaviour is not observed for the magnetic budget using the real part of the magnetic CSDs. This seems to indicate there is an induced phase difference in the coupling of magnetic fields to the LHO and LLO detectors, which is different for the different sites. 
This is as expected given the magnetic fields are fully in phase, as can be seen from the coherence in the left panel of Fig. \ref{fig:MagInj_CohCSD}. This is however not the case for the effect on the GW-strain as seen in the coherence in Fig. \ref{fig:MagInj_StrainCoherence}.
As we do not have phase information for the measured coupling functions used for the noise projection this is not captured in the noise projection.

\begin{figure}
    \centering
    \includegraphics[width=0.49\textwidth]{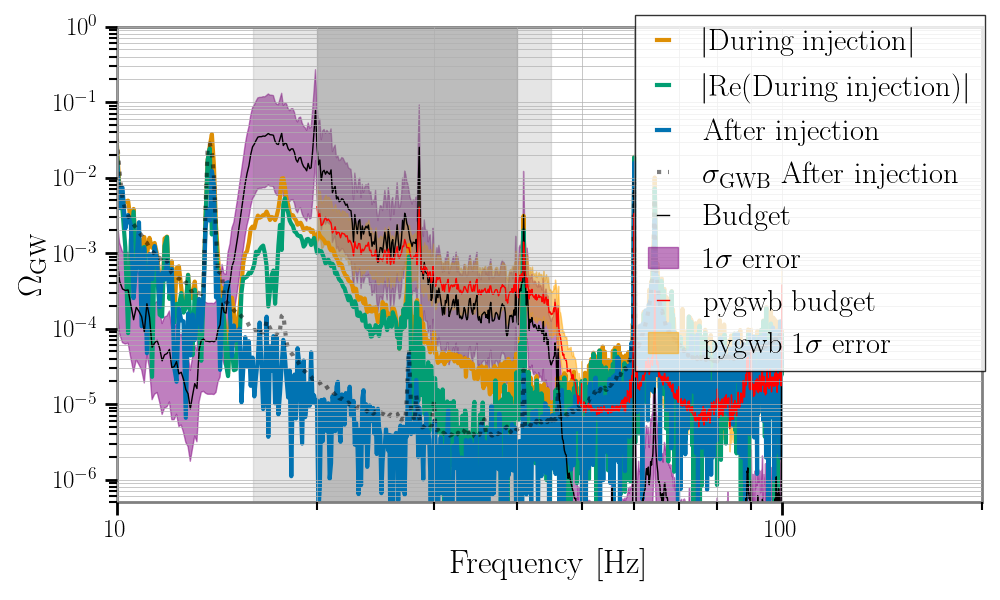}
    \caption{Projection of magnetic noise on the GWB energy density $\Omega_{\mathrm{GW}}$ (black) and its 1-$\sigma$ uncertainty (purple).
    We also show the noise projection using the same {\tt pygwb}-weights applied to the strain measurement (red) and its 1-$\sigma$ uncertainty (orange).
    These noise projections are to be compared with the real value (green) and the absolute value of the complex (yellow) observed cross-correlation statistic. For comparison we also show the absolute value of the cross-correlation statistic from the reference time after the injection (blue) and its $\sigma$ (grey dotted). }
    \label{fig:MagInj_OmegaBudget}
\end{figure}

In Fig. \ref{fig:MagInj_OmegaBudget} we present in black the effect of the correlated magnetic noise injection on $\Omega_{\mathrm{GW}}$, the energy density of GWs that we try to measure. In line with the discussion above, we find that the quadrature sum of the magnetic witness sensors does provide a conservative (over)estimate. Additionally, the failure to capture the phase behavior of the magnetic coupling functions introduces an additional overestimation of the noise projection with respect to the real part of the observed energy GW density.
This highlights the need for future research to investigate the phase behaviour of magnetic coupling functions. Among others it would be of interest to understand the dependence of this induced phase on a number of parameters such as detector site, time and exact coupling location. 

In Fig. \ref{fig:MagInj_OmegaBudget} we also include an alternative calculation of $\Omega_{\mathrm{GW}}$, indicated as the 'pygwb budget' in red. For this calculation we analysed our magnetic data in the same way an by using the same pipeline (called {\tt pygwb}) as our strain data to mimic an as realistic scenario as possible. This is in line with the approach followed by the LVK collaborations when they presnted their results for teh GWB from the first three observing runs \cite{PhysRevD.104.022004}.
more concretly, this pipeline does introduce some slightly different pre- and post-processing of the data compared to the straightforward PSD and CSD calculations performed above. For comparison, we analysed the magnetic data from the injection using the current LVK GWB-analysis pipeline: {\tt pygwb} \cite{Renzini:2023qtj} (more details on {\tt pygwb} can be found in the next section). These results (starting from 20Hz) are presented in Fig. \ref{fig:MagInj_OmegaBudget} in red.
Whereas there are some minor differences between this noise projection and the one shown in black, they are all minor and both of them overestimate the noise contribution. The most noticeable difference is that the error band is smaller for the budget using {\tt pygwb}-weights. However, based on this comparison one would not necessarily prefer one method over the other.

%%%%%%%%%%%%%
\section{Recovery by the search for an isotropic GWB}
\label{sec:Isotropic}
%%%%%%%%%%%%%

The isotropic stochastic search assumes that the GW fractional energy density spectrum, given by Eq. \ref{eq:omegaGW}, can be well described by a simple power-law \cite{PhysRevD.59.102001,Romano:2016dpx}. Under this assumption, the spectrum can be defined by an amplitude $\Omega_{\mathrm{ref}}$ at some chosen reference frequency, $f_{\mathrm{ref}}$, and a spectral index $\alpha$.
\begin{equation}
    \Omega_{\mathrm{GW}}(f)=\Omega_{\mathrm{ref}}\left(\frac{f}{f_{\mathrm{ref}}}\right)^\alpha
\end{equation}

\begin{figure}
    \centering
    \includegraphics[width=\linewidth]{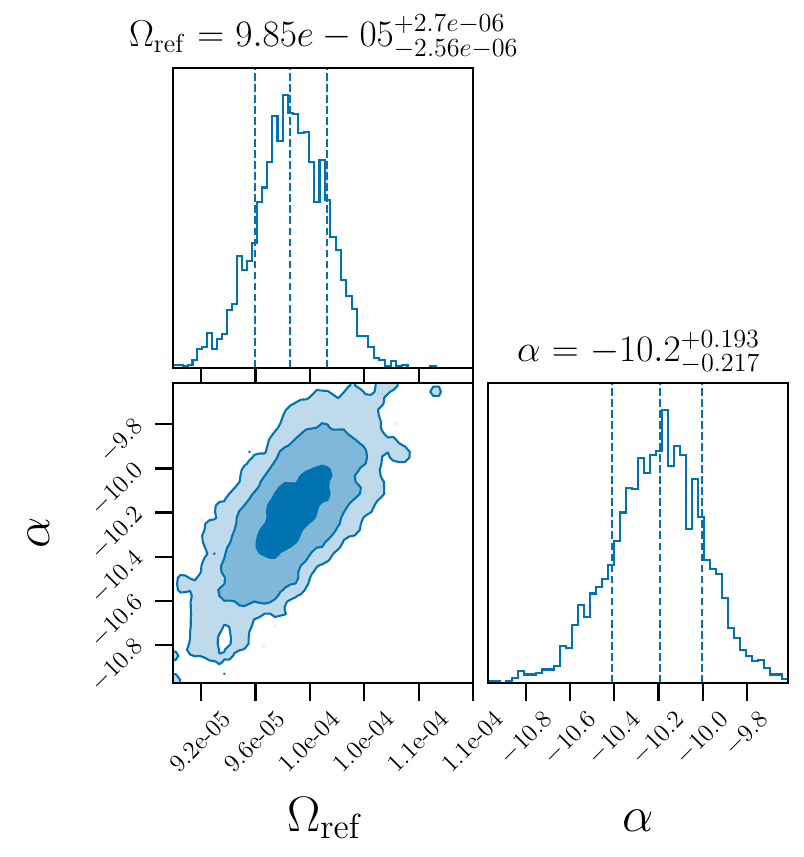}
    \caption{Parameter estimation results on the magnetic injection using a log-uniform prior on $\Omega_{\rm ref}$ and a Gaussian prior on $\alpha$.}
    \label{fig:pygwb_corner_plot}
\end{figure}

To investigate the effects of the correlated magnetic signal in LVK searches for an isotropic GWB, the {\tt pygwb}-pipeline \cite{Renzini:2023qtj} was run on the data containing the injection, as well as a span of data of the same duration that contained no injection for comparison. This standard analysis produces an estimate of the GWB spectrum $\hat{C}_{IJ}(f)$ with standard deviation $\sigma_{IJ}(f)$, as introduced in Eq. \ref{eq:Cij} and Eq. \ref{eq:sigmaGWB}, respectively. This  GWB spectrum and its standard deviation can be used in a hybrid frequentist-Bayesian approach to obtain posterior distributions for the model parameters. For a power-law model, the analysis infers the GWB energy density amplitude at a reference frequency (taken at $25\ \rm{Hz}$) and the spectral index. Using a log-uniform prior on $\Omega_{\rm ref}$ from $10^{-10}$ to $10^{-2}$ and a Gaussian prior on $\alpha$ with a mean of $0$ and standard deviation of $3.5$, inference was performed on the magnetic injection and the results are shown in Fig. \ref{fig:pygwb_corner_plot}. These priors were based on priors used in past LVK analyses for an isotropic GWB \cite{PhysRevD.104.022004}, with the upper boundary on  $\Omega_{\rm ref}$  extended  a few orders of magnitude to ensure the projection of magnetic noise on the GWB energy density is captured.
To further validate our findings, we conducted the same analysis on time-shifted data, which produced the expected outcome of not detecting a signal.

\begin{table}[!htbp]
  \begin{tabular}{|c|c|c|c|c|}
    \toprule
    \hline
    \multirow{2}{*}{\hfill} &
      \multicolumn{2}{c|}{$\alpha = 0$} &
      \multicolumn{2}{c|}{$\alpha = -10.2$} \\
      \hline
      & No injection & Injection & No injection & Injection \\
      \hline
      \midrule
    $\ \hat{C}^{(\alpha)}_{\rm ref}\ $ & $3.51 \times 10^{-7} $ &  $4.43\times 10^{-6}$ & $9.31\times 10^{-7}$ & $9.87\times 10^{-5}$  \\
    $\ {\sigma}^{(\alpha)}_{\rm ref}\ $ & $3.93\times 10^{-7}$ & $4.88\times 10^{-7}$ & $5.59\times 10^{-7}$ & $1.87\times 10^{-6}$  \\
    $\ \rho^{(\alpha)}_{\rm ref}\ $ & $0.89$ & $11.26$ & $1.67$ & $176.57$  \\
    \hline
    \bottomrule
  \end{tabular}
\caption{Point estimates, standard deviations, and signal-to-noise ratios for spectral indices $0$ and $-10.2$ for data containing the injection as well as a stretch of data of equal duration which did not contain any injection. As explained in text, we use ${\sigma}^{(\alpha)}_{\rm ref} $ from   the `no injection' time for the calculation of all signal-to-noise ratios.}
\label{table:pygwbResults}
\end{table}

The frequentist analysis produces a point estimate and variance at a reference frequency for spectral index $\alpha = 0$. These results are then reweighted for other values of $\alpha$ \cite{PhysRevD.104.022004}. We report the point estimate, standard deviation and signal-to-noise ratio for both $\alpha=0$ and $\alpha=-10.2$, the inferred value of $\alpha$ from the parameter estimation. The signal-to-noise ratio, $\rho^{(\alpha)}_{\rm ref}$, is computed as the ratio of the point estimate and standard deviation
\begin{equation}
    \rho^{(\alpha)}_{\rm ref} = \frac{\hat{C}^{(\alpha)}_{\mathrm{ref}}}{\sigma^{(\alpha)}_{\mathrm{ref}}},
\end{equation}
where the standard deviation is always taken from the data which did not contain the injection, as a loud signal biases its value \cite{PhysRevD.59.102001}. The results are shown in Table \ref{table:pygwbResults}.

The results of the search for an isotropic GWB are consistent with the estimated effect on $\Omega_{\mathrm{GW}}$ shown in Fig. \ref{fig:MagInj_OmegaBudget}.

%%%%%%%%%%%%%
\section{Noise subtraction: showcase}
\label{sec:NoiseSubtraction}
%%%%%%%%%%%%%

Previous work, such as \cite{Thrane:2014yza,Coughlin:2016vor,Coughlin:2018tjc}, has looked into using Wiener filters to subtract magnetic noise from GW strain measurements in case correlated magnetic noise would couple significantly. However, all of these studies either relied on simulated data or used different magnetometers as proxies for a contaminated GW strain channel. With the current data, we are uniquely positioned, for the first time, to use actual strain data with correlated broadband noise present and subtract it by relying on witness sensors (magnetometers) located at the detector sites.
Despite this unique testing scenario, future work would have to investigate in more detail to what extent the findings presented in this section are affected by the large magnitude of the correlated noise signal present in the dataset. It is highly likely that more realistic scenarios with lower amplitude correlated noise contributions would require more advanced techniques.

\subsection{Noise subtracted data}

For this demonstration we used the entire noise dataset to train our Wiener filter using 100 second windows. We always used a single witness sensor for noise subtraction, but tested a number of different combinations, such as the orientation of the witness sensor as well as its location. In what follows we will focus on noise subtraction using the $x$-direction magnetometers located at `Vertex' as the witness sensor. As a short summary, the horizontal sensors perform better compared to the vertical magnetometers for noise subtraction. This is in line with the smaller response in the vertical magnetometers. Secondly, there is no significant difference between the different sets of magnetometers throughout the building. Finally, noise subtraction at LHO (LLO respectively) with witness sensors from the other detector site, i.e. LLO (LHO respectively) works decently, but does not perform as well as using witness sensors at the same detector site. 

We can think of the strain data of interferometer $I$ as a sum of a GW component $h_I$, a detector noise component $n_I$ and a magnetic noise component $m_I$, which couples through the coupling function $\Kappa_I$ as discussed earlier. This yields,
\begin{equation}
    \label{eq:Wiener_data}
    s_I = h_I + n_I + \Kappa_I  m_I        
\end{equation}
where $\langle n_I n_J\rangle=0$ if $I\neq J$. Rather than using $\Kappa_I$ measured during the injections, we will use a Wiener filter to estimate this quantity in a more accurate way. To this extent we use $w_I$ to refer to the measured data from the witness sensor (magnetometer), which contains a measurement of $m_I$ as well as  a different independent noise component $\eta_I$,
\begin{equation}
    \label{eq:Wiener_data_witness}
    w_I = \eta_I + m_I.        
\end{equation}

Some examples of $\eta_I$ are local magnetic fields from e.g. electronic devices operating in the vicinity and intrinsic sensor noise.
The Wiener filter for the magnetic coupling function, $\Kappa_{I,Wiener}$, can be constructed as,
\begin{equation}
    \label{eq:Wiener_filter}
   \Kappa_{I,Wiener} = \frac{\langle s_I w_I\rangle}{\langle w_I w_I\rangle}.        
\end{equation}
Due to the large dynamic range of $s_I$ we first apply a high-pass filter. If one does not perform this step the very large strain noise at low frequencies (e.g. $10^{-18}$ near 6Hz) will prevent the Wiener filter from being sensitive to anything   with an amplitude of about $10^{-22}$ in the region of interest. In our case we use an 8th-order high-pass Butterworth filter with a cutoff frequency of 16Hz.
Additionally, due to the significant computation time, the data was down-sampled to 128Hz to only focus on the sub-50Hz region where the injection occurred. 

\begin{figure*}
    \centering
    \includegraphics[width=0.49\textwidth]{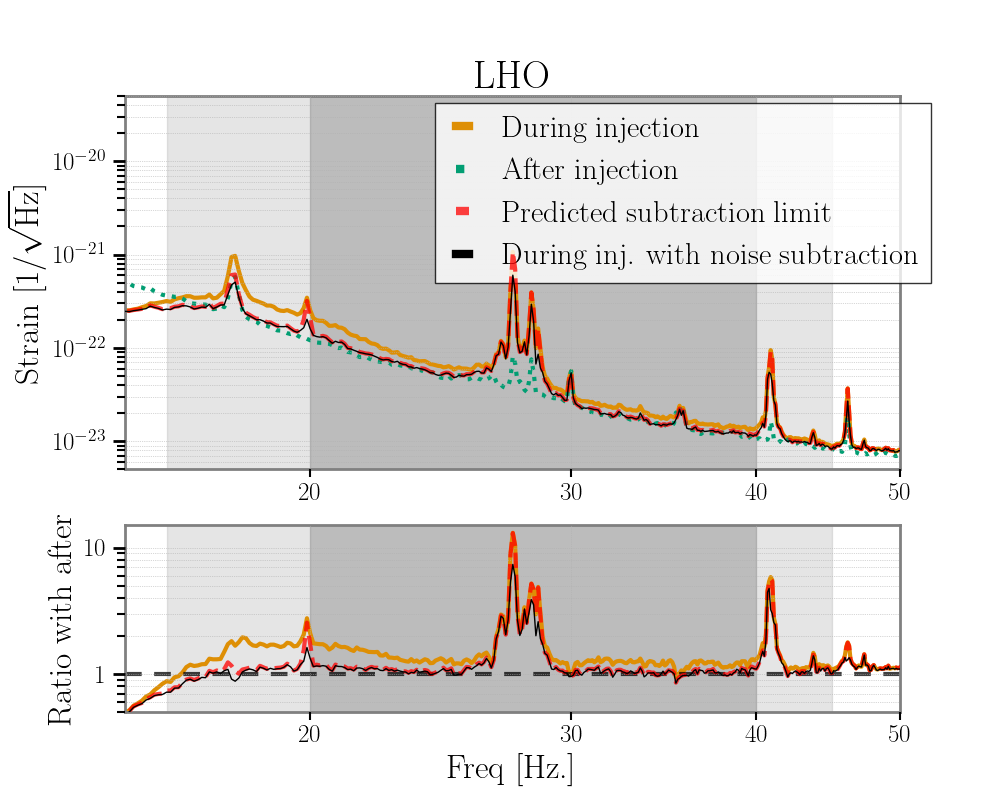}
    \includegraphics[width=0.49\textwidth]{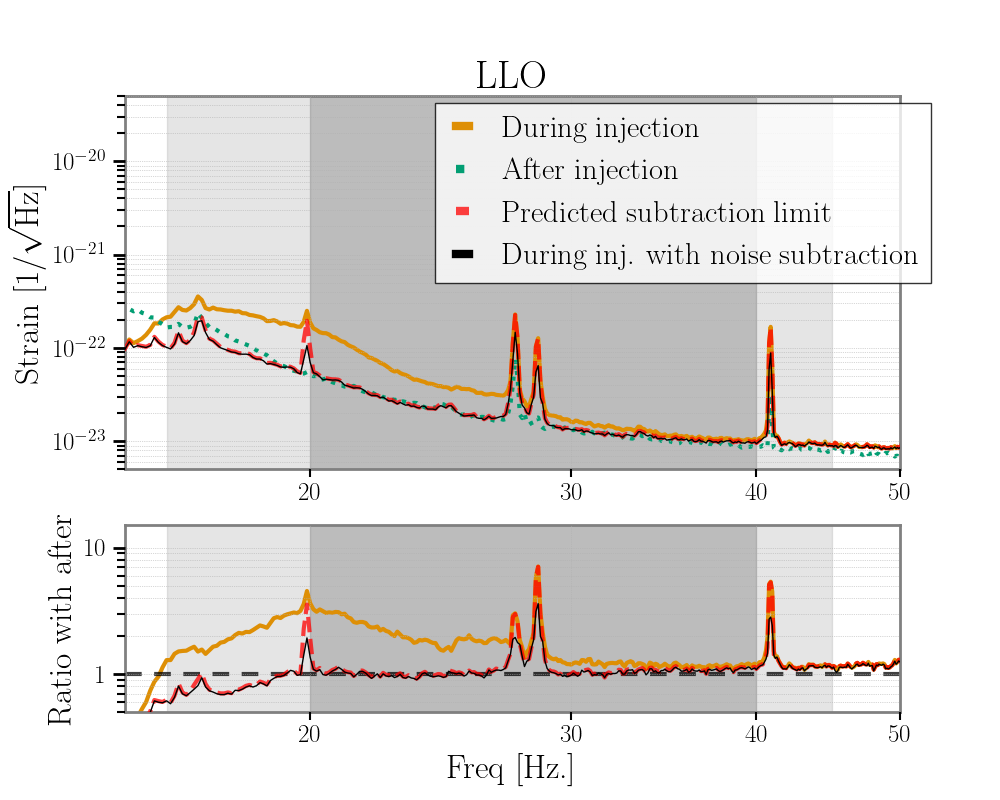}
    \caption{The top panels show the ASD strain for LHO (left) and LLO (right). The bottom panels represent the ratio of ASDs with respect to the reference strain measurement after the injection. Data taken during the injection is shown in yellow, whereas the green dashed line represents the reference strain measurement after the injection in the top panels. Red shows the predicted subtraction limit achievable, whereas in black we show the actual subtraction achieved.
}
    \label{fig:ASD_subtraction}
\end{figure*}

Figure \ref{fig:ASD_subtraction} shows the strain ASD for LHO and LLO and compares the data during the injection both without (yellow) and with (black) noise subtraction to the data observed during the reference time after the injection (green dashed). In red we also include the level of noise subtraction which is theoretically possible based on the coherence between the target and witness channel for Wiener filtering. We conclude the time-based Wiener filter using one witness channel is able to reach, and in some frequency regions even outperform, its predicted subtraction limit based on the coherence between witness and target channels. Apart from some narrowband effects and the cluster of peaks below 30Hz in both sites (pre-dominantly LHO), the subtracted data does agree closely to the reference data after the injection. As mentioned earlier, some of these are associated with suspension resonances triggered by the magnetic injection. Additional witness sensors could potentially further reduce their effect. However, this is considered to be outside the scope of this work.

\begin{figure*}
    \centering
    \includegraphics[width=0.49\textwidth]{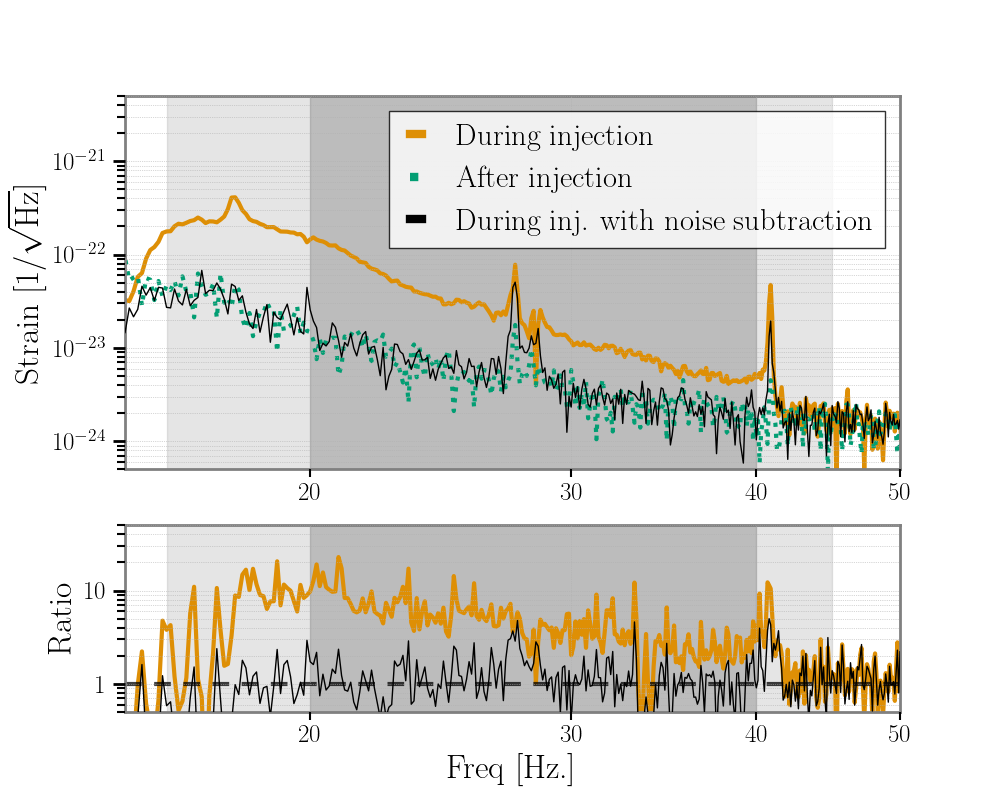}
    \includegraphics[width=0.49\textwidth]{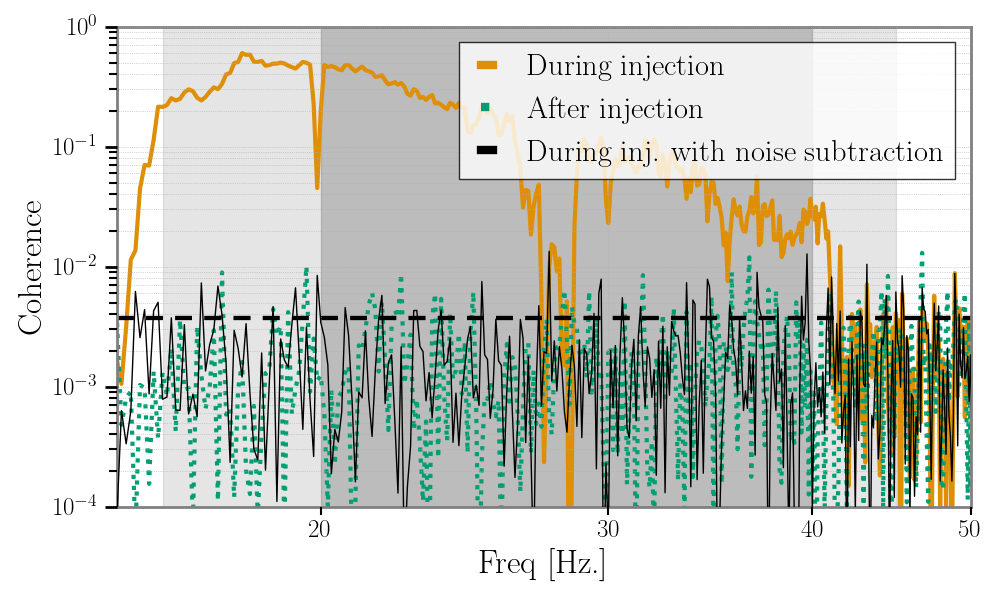}
    \caption{The left panels show the strain CSD between LHO and LLO (top) during injection (yellow), after injection (green dashed) as well as the injection data with noise subtraction applied (black). The bottom panel shows the ratio with respect to a reference measurement after the injection.
    The right panel shows the coherence between LHO and LLO for the same datasets.}
    \label{fig:CSD_Coh_subtraction}
\end{figure*}

Additionally we present the strain CSD between LHO and LLO in the left panel of Fig. \ref{fig:CSD_Coh_subtraction} and its coherence in the right panel of the same figure. We get excellent agreement between the noise subtracted data and the reference measurement after the injection and find no significant coherence for either.

We demonstrated that with a relatively simple Wiener subtraction relying only on one witness sensor at a single site, we are able to achieve sufficient levels of noise subtraction. Here we need to add the caveat that this noise subtraction is performed in a scenario where the injected signal is so loud that the signal-to-noise ratio of the magnetic field in the magnetometer is very large. Also for the strain channel, the magnetic noise forms a significant contribution to the ASD during the time of the injection. In a more realistic scenario this will not be the case. The magnetic signal-to-noise ratio will still be significant in the witness sensor, but by no means as large as in this specific dataset. Additionally, in a more realistic scenario, the magnetic noise will be below the ASD noise floor for  most of its frequency range, where only future third-generation detectors such as the Einstein Telescope and Cosmic Explorer could have more significant magnetic noise coupling. Both of these features  mean that the noise subtraction might behave differently in a more realistic scenario. Hence, we strongly recommend for future follow-up studies additional correlated magnetic injections with injection strengths that are below the ASD noise floor but couple significantly to the CSD spectra over the integrated time of the injection. This will more closely mimic a realistic situation for future analyses when environmental noise starts to couple coherently to the detectors.
Secondly, we discussed the importance of using high-pass filters as the observed strain spans many orders of magnitude. This increases the risk of introducing noise components. Realistically, future efforts might need to focus on multiple smaller frequency regions in which the power is about the same order of magnitude to achieve optimal noise subtraction, after which the different analyses across the different frequency bands are to be combined.
Finally due to the computational cost to calculate precise Wiener filters over thousands of seconds, we downsampled our dataset to 128Hz, corresponding to a Nyquist frequency of 64Hz, yielding a single-core computation time of $\mathcal{O}$(20-60 min) for the subtraction of noise in one target channel using one witness channel. This indicates that it might be very computationally costly and potentially impossible to perform noise subtraction on $\sim$1yr of data in the case of higher-frequency correlated noise\footnote{Currently Wiener filtering is succesfully used by the LVK Collaboration to remove noise features. However, the Wiener filters are typically trained on tens of seconds of data which then afterwards are applied over longer stretches of data. For extremely low coherent noise sources it might be needed to perform this training on months of data at the same time, which does lead to significantly increased computational cost, which is not an issue for other use cases.}. Conveniently, most of the predicted broadband correlated noise sources are predicted to be situated in the lower frequency region of Earth-based GW interferometers.

\subsection{GWB signal recovery on noise subtracted data}
We perform the same studies as in Sec. \ref{sec:Isotropic} on the noise subtracted data, including both timeshift and zero-lag analyses. The frequentist analysis results for the point estimates, standard deviations, and signal-to-noise ratios are shown in Table \ref{table:pygwbNoiseSub} for the same spectral indices considered previously. The consistency of the standard deviations in the timeshift and zero-lag cases for each spectral index indicates that the noise subtraction was successful. Additionally, the signal-to-noise ratios are significantly lower compared to the data before applying Wiener filtering. However, the absolute value  of the signal-to-noise ratios for both $\alpha=0$ and $\alpha=-10.2$ are larger than one typically would expect from noise-only scenarios.

\begin{table}[!htbp]
  \begin{tabular}{|c|c|c|c|c|}
    \toprule
    \hline
    \multirow{2}{*}{\hfill} &
      \multicolumn{2}{c|}{$\alpha = 0$} &
      \multicolumn{2}{c|}{$\alpha = -10.2$} \\
      \hline
      & Timeshift & Zero-lag & Timeshift & Zero-lag \\
      \hline
      \midrule
    $\ \hat{C}^{(\alpha)}_{\rm ref}\ $ & $-9.97\times 10^{-7}$ &  $1.00\times 10^{-6}$ &  $-1.84\times 10^{-6}$ & $-3.22\times 10^{-6}$  \\
    $\ {\sigma}^{(\alpha)}_{\rm ref}\ $ & $3.45\times 10^{-7}$ & $3.45\times 10^{-7}$ & $4.42\times 10^{-7}$ & $4.40\times 10^{-7}$  \\
    $\ \rho^{(\alpha)}_{\rm ref}\ $ & $-2.89$ & $2.90$ & $-4.16$ & $-7.32$  \\
    \hline
    \bottomrule
  \end{tabular}
\caption{Point estimates, standard deviations, and signal-to-noise ratios using spectral indices $0$ and $-10.2$ for timeshift and zero-lag analyses on the noise-subtracted data.}
\label{table:pygwbNoiseSub}
\end{table}

\begin{figure*}
    \centering
    \includegraphics[width=.49\textwidth]{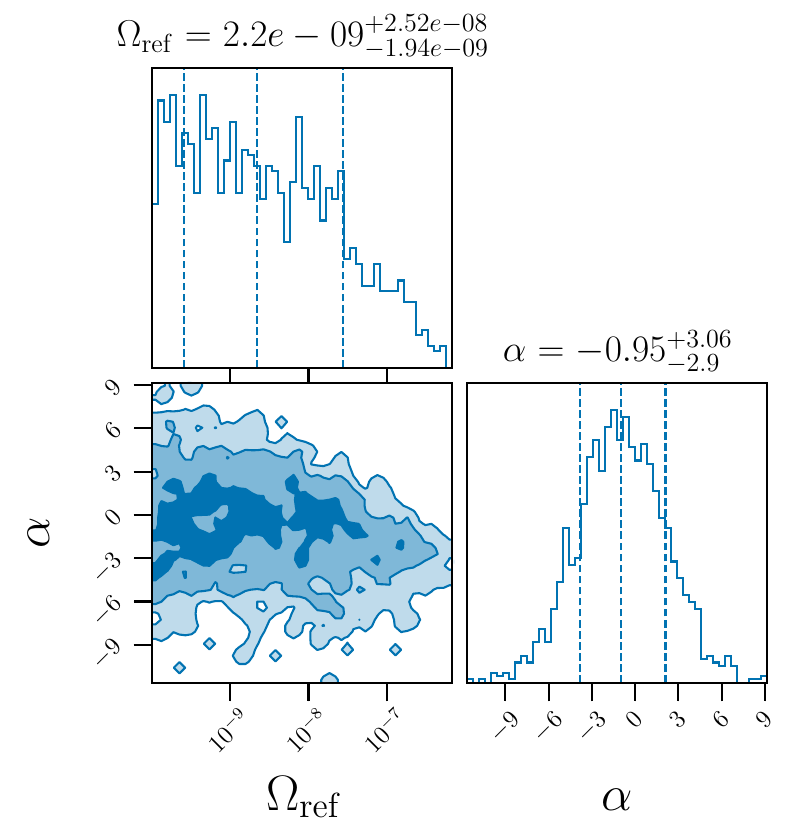}
    \includegraphics[width=.49\textwidth]{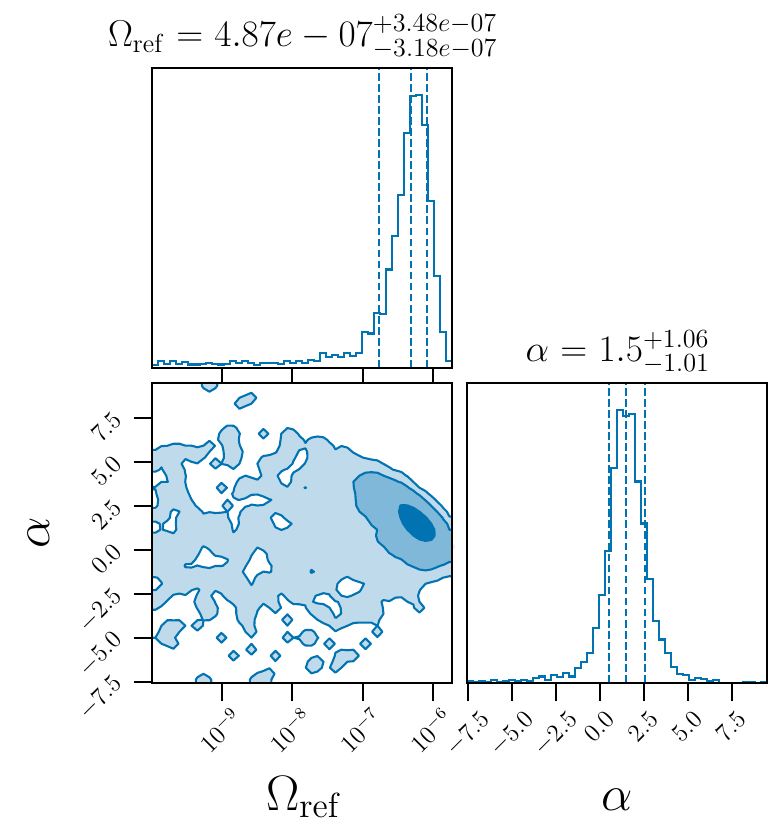}
    \caption{Parameter estimation on time-shifted (left) and zero-lag (right) noise-subtracted data.}
    \label{fig:pygwb_noise_sub}
\end{figure*}

We also ran parameter estimation on both the time-shifted and zero-lag data. Using the same priors described in Sec. \ref{sec:Isotropic}, we find that the posteriors in the timeshift analysis are consistent with the priors and do not recover a signal. However, the zero-lag results reveal a hint of a signal with a spectral index around $1.5$. Both corner plots are shown in Figure \ref{fig:pygwb_noise_sub}.

Reweighting the frequentist results to $\alpha=1.5$, yields a point estimate and standard deviation of $6.61\times 10^{-7}$ and $1.91\times 10^{-7}$, respectively. These values correspond to a signal-to-noise ratio of $3.45$. Given that the point estimate is consistent with zero at the $3-4\sigma$ level, this does not constitute a detection of some signal with a spectral index of $1.5$, for which typically a value of $5\sigma$ is required. However, anything $>3\sigma$ would typically be considered as strong evidence for a potential signal. 
In combination with the relatively large signal-to-noise ratios we observed for the other spectral indices, we believe this indicates that the subtraction of the magnetic noise may lead to non-trivial effects on the data.
Given that the coherent magnetic injections were very strong, this problem may not occur with respect to the subtraction of the real Schumann resonance produced signals.
One potential cause is that even after the applied high-pass filter, the dynamic range of the observed strain was too large and this left a small to unnoticeable effect on the strain data at higher frequencies which shows up in the more sensitive GWB analysis.
These effects should be further investigated in future efforts, but are outside the scope of this paper.

%%%%%%%%%%%%%
\section{Conclusion}
\label{sec:Conclusion}
%%%%%%%%%%%%%

In this paper, we presented the results of the first-ever, broadband correlated magnetic noise injection between two Earth-based GW detectors, LIGO Hanford and LIGO Livingston, which are separated by $\sim$3000km. We demonstrated that the injection was successfully able to create coherent magnetic noise between magnetometers at both sites as well as in the strain measurement by simultaneously injecting magnetic noise with the same predefined phase.
Afterwards we showed that the observed effect in the strain channels agrees well with the projected effect based on the observed tri-axial magnetic noise and the magnetic coupling function measured during the O4a observing run. The noise projection for the correlated strain measurement, however, does show an overestimation of more than 1$\sigma$ across the entire frequency band of the injection. We highlight the likely key origins of this (slight) overestimation. First of all, it seems the quadrature sum of all nine combination of the tri-axial magnetic measurements at each site cause an overestimation. Secondly, the imaginary component of the strain correlation is non-negligible, particularly above 30Hz. However, at the same time this is not the case for the magnetic noise. This leads us to conclude the magnetic noise coupling at both sites does have a different phase partially de-correlating our injected signal. In the past, only the magnitude of the magnetic coupling has been reported. However, this work suggests that the phase might have a non-negligible impact when looking at cross-correlations. 

Apart from demonstrating the strength and weaknesses of the noise projections we also analysed the data using the isotropic GWB search pipeline {\tt pygwb}. The optimal power-law signal we recover has a slope $\alpha=-10.20^{+0.19}_{-0.22}$ and an amplitude $\Omega_{\rm ref} = 9.85^{+0.27}_{-0.26}\times 10^{-5}$. If we fix the spectral index to the recovered value, we find a signal-to-noise ratio of about 50. However, this value includes uncertainty that comes from the loudness of the injected magnetic signal. If we use the uncertainty based on a quiet time without noise injection the significance increases to a signal-to-noise ratio of 176.57. This is in line with the earlier projections from the budget. We would like to highlight that the steepness of this signal is a benefit for future separation of magnetic noise and a GWB signal. First of all, such a large negative slope is not expected from actual physical signals. Secondly, this large discrepancy in spectral index of the magnetic noise and GWB signal indicates that a joint parameter estimation and spectral separation effort could be very successful \cite{PhysRevD.102.102005}.

Finally, we applied a simple time-domain Wiener filter using one of the magnetometers as a witness sensor to subtract the noise from the strain channel. In this proof-of-principle study we achieved excellent noise subtraction and recovered a strain ASD close to the observed value in the quiet reference time after the injection. Also, the correlated noise in the strain CSD and coherence was entirely removed.
To achieve this subtraction we down-sampled the data to 128Hz to achieve reasonable computation times for this 43 min-long data segment. Additionally we applied an 8th order Butterworth high-pass filter to decrease the dynamic range of the strain amplitude to be sensitive to the smaller amplitudes in the frequency range of interest.
Despite the (apparent) success of noise subtraction based on the ASD and CSD, several questions remain. Namely, if we perform the search for an isotropic GWB on this dataset, we find marginal (3.45 $\sigma$) evidence for a power-law with a spectral index of $\alpha=1.5$. Additionally, we find a large negative signal-to-noise ratio of $-7.32$ for the spectral index present in the un-subtracted dataset, $\alpha=-10.2$. This indicates that while the Wiener filter is able to successfully subtract the noise, it might alter the data in unknown and unforeseen ways impacting our search. We believe caution is needed for future efforts and additional investigations are needed.

The injection we analyzed in this work provided many important insights in correlated noise coupling to a GWB search and how one can hope to mitigate its effect. However, it also provides many opportunities for more in depth follow-up studies, which we will highlight below. First of all, we believe a reduced amplitude signal (injected for a potentially longer period) would be good to simulate a more real-life case where the noise is subthreshold, but still can be recovered with a    5-10$\sigma$ significance. Additionally if one performs a significantly weaker injection over a period of $\sim 1$ day (either continuous or in segments), this could provide interesting insights on the potential impact of correlated noise for the search for an anisotropic GWB as well as searches for continuous GWs \cite{PhysRevD.111.082005}.
Finally, a joint injection of a  GWB signal and correlated noise having comparable amplitudes would provide many data analysis opportunities. It would be the ideal demonstrator case for spectral separation in parameter estimation. Furthermore, it would allow one to investigate potential biases introduced from noise subtraction on signal recovery.
With the LIGO, Virgo and KAGRA collaborations reaching the needed sensitivity to observe a GWB signal in the next years to a decade, now is the ideal time to further investigate these open questions and prepare mitigation strategies and techniques for correlated noise.

\acknowledgements

K.J. was supported by FWO-Vlaanderen via grant number 11C5720N during part of this work. J.L. was supported by NSF Award PHY-2207270.
M.W.C acknowledges support from the National Science Foundation with grant numbers PHY-2308862 and PHY-2117997.
This material is based upon work supported by NSF’s LIGO Laboratory which is a major facility fully funded by the National Science Foundation.
The authors are grateful for computational resources provided by the LIGO Laboratory and supported by National Science Foundation Grants PHY-0757058 and PHY-0823459.

%%%%%%%%%%%%%%%%%
\bibliography{references}
%%%%%%%%%%%%%%%%%

\appendix

\section{Inputoptics and Outputoptics results}
\label{appendix:inputOutput}

\begin{figure*}
    \centering
    \includegraphics[width=0.49\textwidth]{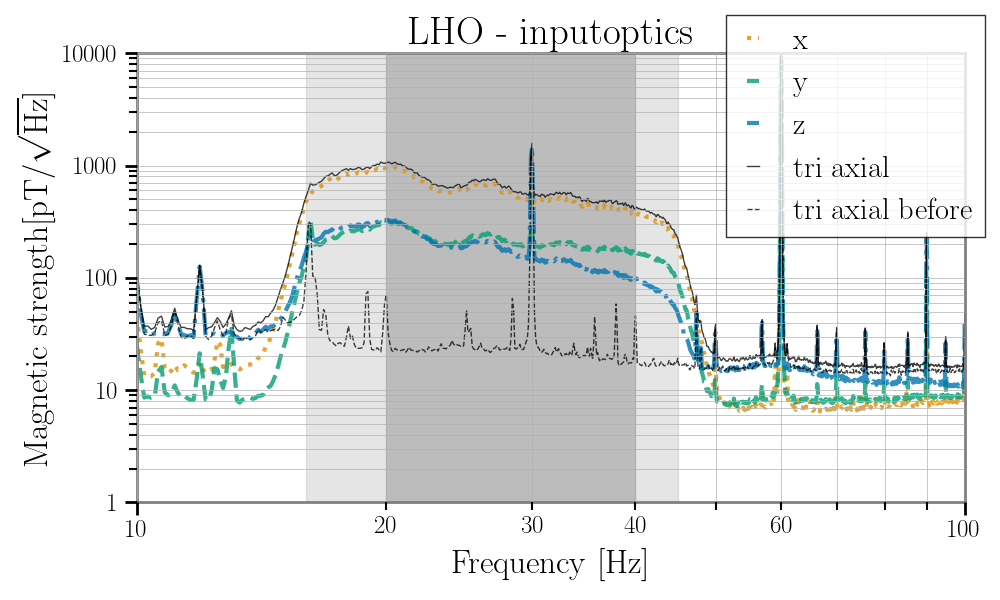}
    \includegraphics[width=0.49\textwidth]{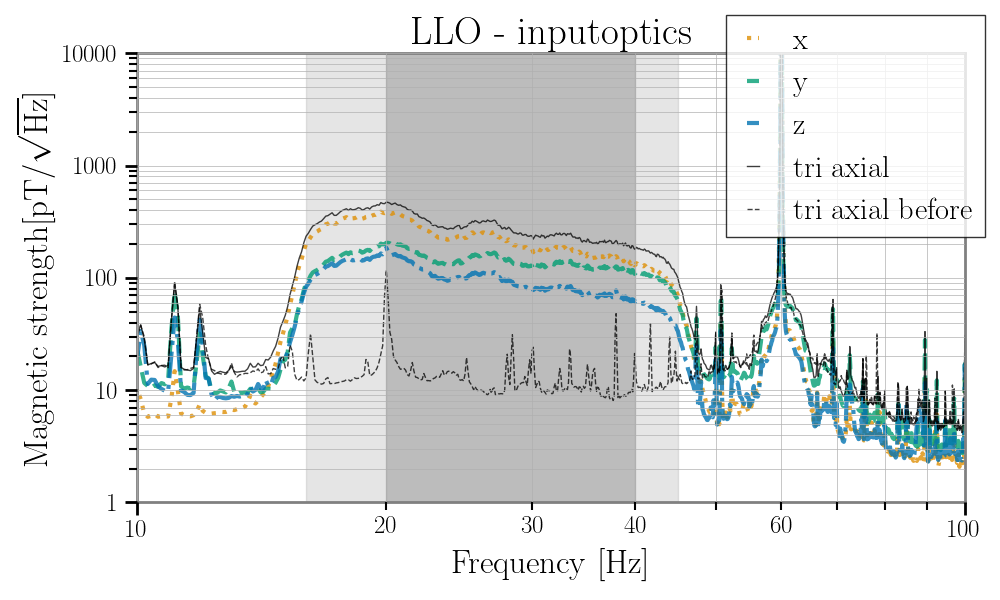}
    \includegraphics[width=0.49\textwidth]{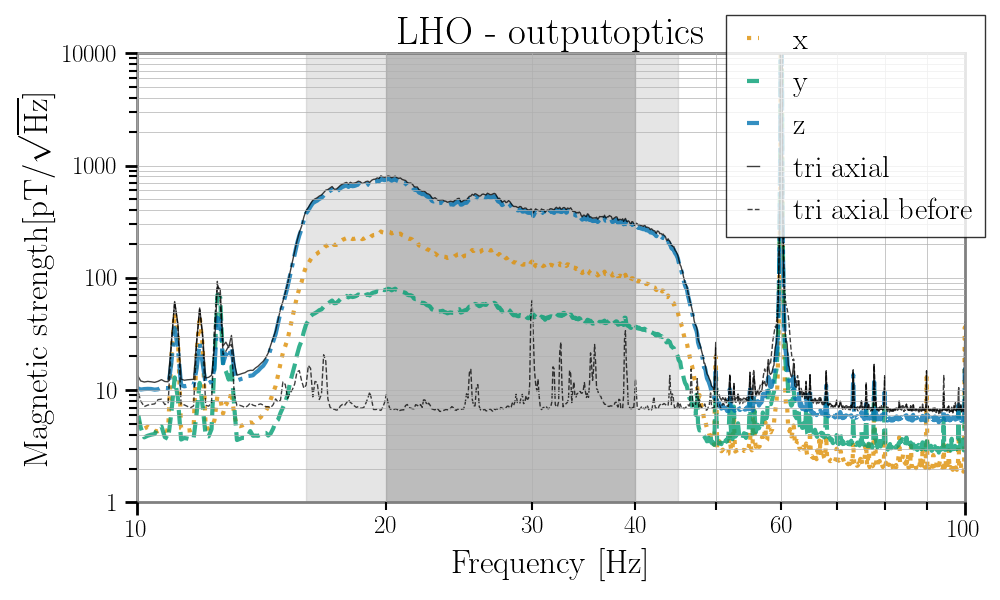}
    \includegraphics[width=0.49\textwidth]{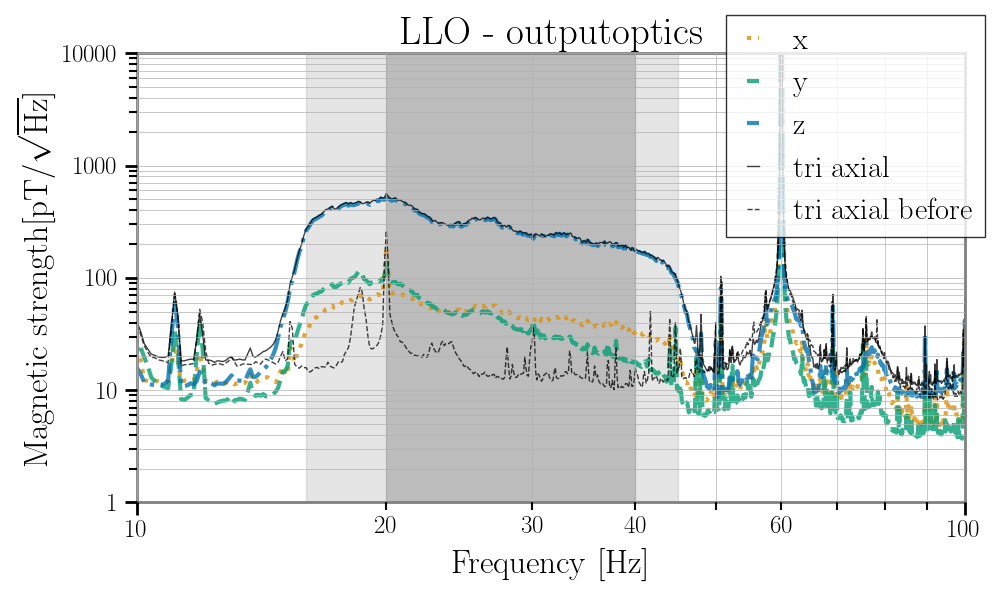}
    \caption{The magnetic amplitude spectral density for LHO (left) and LLO (right) during the injection. The top and bottom panels respectively show the magnetic ASD at the `Inputoptics' and `Outputoptics' locations. For the time of the injection we show the three independent components, x (dotted, yellow), y (dashed, green), z (dot-dashed, blue) as well as tri-axial (solid, black). For comparison we also show the magnetic noise observed during the reference time after the injection (dashed, black). The dark and light gray vertical bands show the frequency range in which the noise was injected, respectively tapered off.}
    \label{fig_appendix:MagInj_ASD}
\end{figure*}

\begin{figure*}
    \centering
    \includegraphics[width=0.49\textwidth]{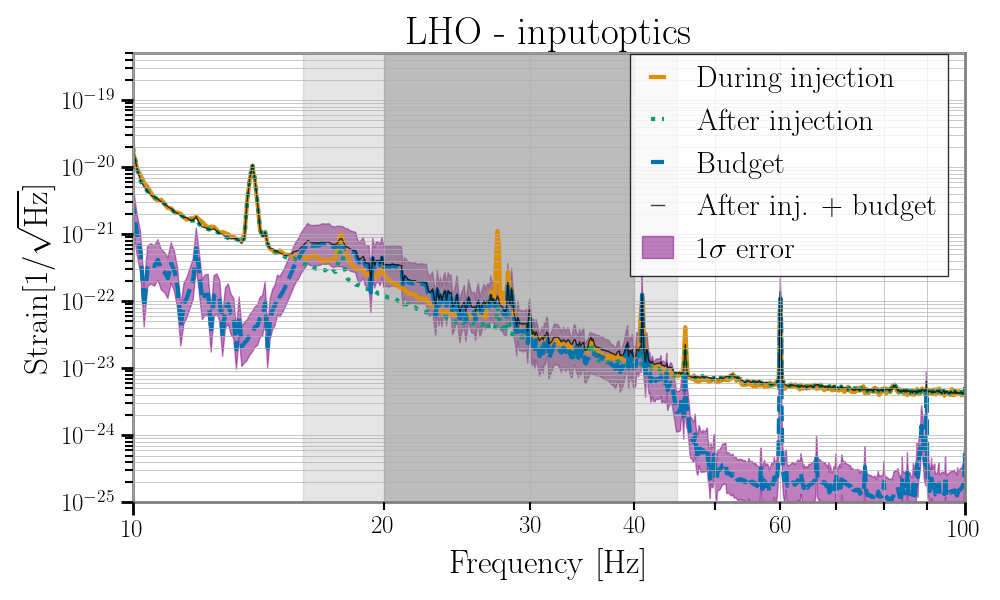}
    \includegraphics[width=0.49\textwidth]{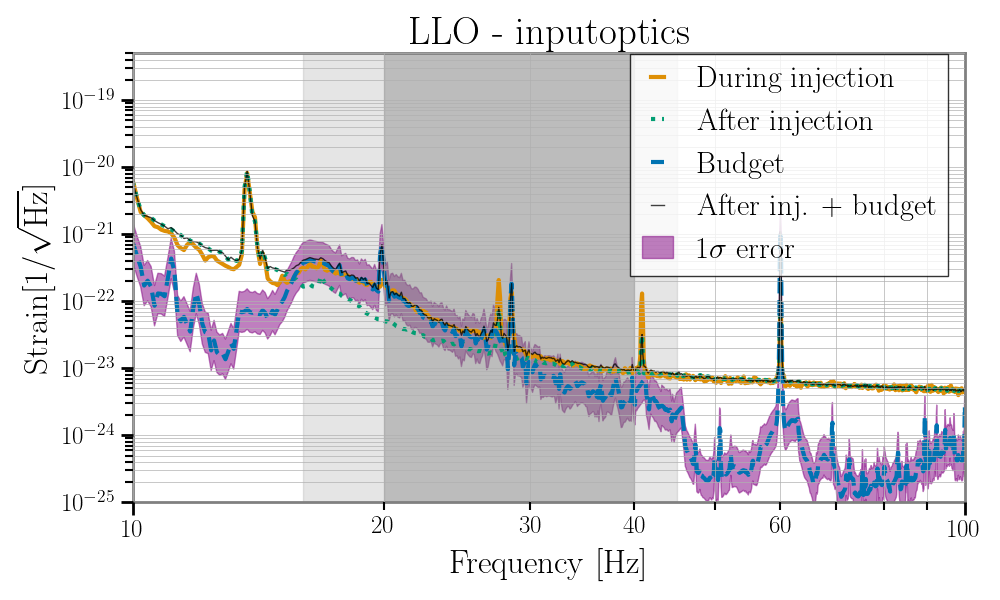}
    \includegraphics[width=0.49\textwidth]{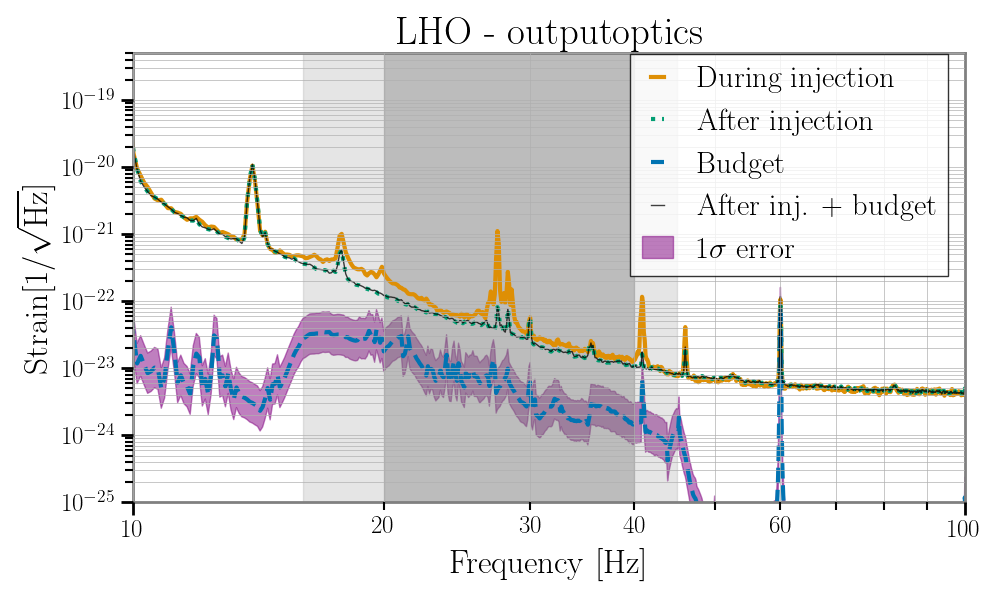}
    \includegraphics[width=0.49\textwidth]{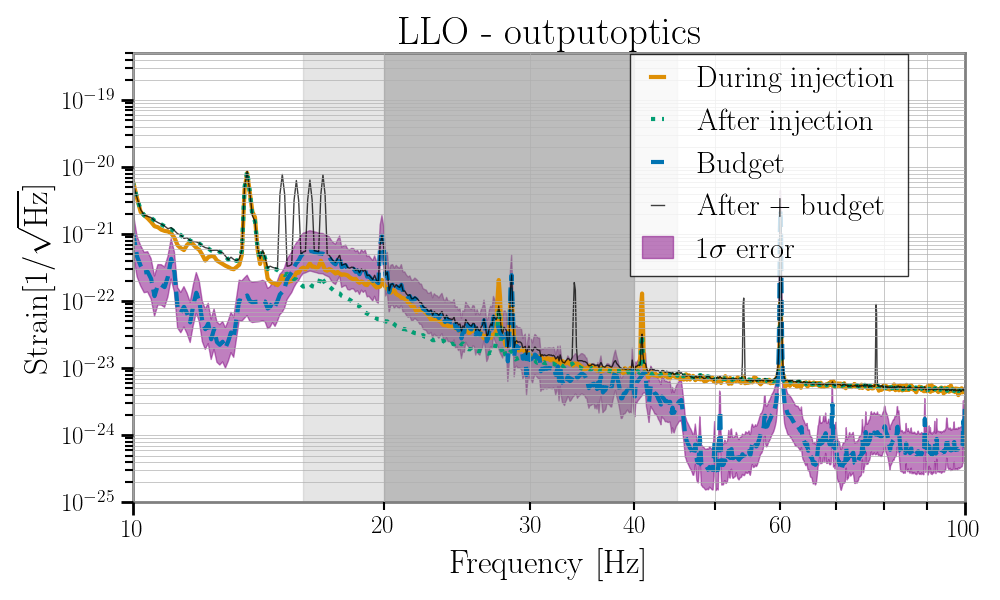}
    \caption{The projections of the magnetic noise on the GW strain ASD of LHO (left) and LLO (right) during the injection. The top and bottom panels respectively show the noise budget at `Inputoptics' and `Outputoptics'. The dark and light gray vertical bands show the frequency range in which the noise was injected, respectively tapered off.}
    \label{fig_appendix:MagInj_ASDBudgets}
\end{figure*}

\end{document}